\documentclass[useAMS,usenatbib]{mn2e}
\usepackage{times}
\usepackage{graphicx}

\begin{document}

\title[CG J1720-67.8: Dynamics and Star Formation]{Dynamics and star formation
activity of CG J1720-67.8 unveiled through integral field spectroscopy and radio 
observations\thanks{Based on data collected at the Anglo-Australian 
Telescope, Siding Spring, Australia (Proposals PATT/02A/24,
PATT/03A/22), at the Magellan Telescope, Las Campanas Observatory, 
Chile, at the Australia Telescope Compact Array, Narrabri, NSW (Proposal C1026), 
and at the ESO 3.6 m telescope (Proposal 63-N-0737).}}

\author[S. Temporin et al.]{Sonia Temporin,$^1$\thanks{giovanna.temporin@uibk.ac.at} 
Lister Staveley-Smith,$^2$\thanks{lister.staveley-smith@csiro.au} 
Florian Kerber$^3$\thanks{fkerber@eso.org}\\
$^1$Institut f\"ur Astrophysik, Leopold-Franzens-Universit\"at Innsbruck, 
Technikerstra\ss e 25, A-6020 Innsbruck, Austria\\
$^2$Australia Telescope National Facility, CSIRO, P.O. Box 76, Epping, NSW 1710, Australia\\
$^3$Space Telescope European Coordinating Facility, European Southern 
Observatory, Karl-Schwarzschild-Stra\ss e 2, D-85748 Garching, Germany}

\maketitle
\begin{abstract}
CG J1720-67.8 is an ultra compact group of several galaxies with a low
velocity dispersion, and displaying the hallmarks of mutual
interaction and possible tidal dwarf galaxy formation. In hierarchical
models, the system is a possible precursor to a massive elliptical
galaxy.  In this paper, we use new optical integral field
spectroscopic and radio observations to investigate the evolutionary
status of the group in more detail: global star-formation rates are estimated using
H$\alpha$ and 1.4 GHz radio continuum measurements; H\,{\sc i}
observations provide an upper limit to the global neutral gas content;
optical broadband colours and spectra provide ages and stellar mass
estimates for the tidal dwarf candidates; the bidimensional H$\alpha$
velocity field is used to trace the kinematics of the group and its members,
which are compared with numerical simulations of galaxy
encounters. The observations suggest a model in which multiple
interactions have occurred, with the latest strong encounter involving
at least two major components within the last 200 Myr. Debris from the
encounter fuels ongoing star formation at the global level of $\sim20$
M$_{\odot}$ yr$^{-1}$, with self-gravity within the tidal tail possibly
providing a mechanism to enhance the star formation rate of the tidal
dwarf candidates, with bursts of star-formation in clumps of mass
$\sim2\times 10^7$ M$_{\odot}$ appearing within the last 10 Myr. 
The amount of time required for final merging of all group
components remains uncertain.
\end{abstract}


\section{INTRODUCTION}

One of the most important problems in cosmology concerns the formation
and evolution of galaxies. Sixty per cent of galaxies are found in
groups \citep{tu87}. This fact alone indicates that the group
environment is of utmost importance for galaxy evolution. The most
recent views of hierarchical structure formation in the Universe
involve galaxy groups as building blocks of clusters and as places
where interactions pre-process galaxies, which eventually become part
of the cluster population \citep[e.g. ][]{mi04}.
Due to the high densities and low velocity dispersions
(with respect to clusters) that characterize them, compact groups (CGs) of
galaxies offer
the most favorable environmental conditions for interaction and merging processes to
take place and lead to profound galaxy transformations. 
The evolution of CGs 
is believed to end in bright elliptical galaxy formation through subsequent mergers \citep[e.g., ][]{kc98,ba98}. 
Therefore, groups
can be regarded as putative factories of field elliptical galaxies. This evolutionary
scenario is supported by the recent identification of ``fossil groups'' or
``over-luminous elliptical galaxies'' in the X-ray domain. More than 10 fossil groups
are known to date, although only a few of them have been studied in detail 
\citep[see e.g., ][and references therein]{s04}. The most likely
interpretation of these fossil systems is that they are the end-results of galaxy
merging within normal groups \citep{j03}.

However, despite considerable efforts, details of the processes leading from CGs to
elliptical galaxies are still uncertain and, particularly, the latest stages of
evolution of CGs are profoundly unknown. In fact only very few high density (i.e.
evolved) CGs have been found to date. The most well studied sample of CGs
is that compiled by \citet{hi82} which encompasses groups in
differing evolutionary stages. However, none of these systems can be said to be close
to final merging, with the possible exception of HCG 31 \citep{ler04} and HCG 79
which is estimated to achieve final coalescence in less than 1 Gyr \citep{ni00}.
The rarity of highly evolved CGs suggests that, once triggered, the coalescence
phase must be relatively short. On the other hand, the fraction of fossil groups with
respect to normal groups would point to a relatively slow rate of formation of fossil
groups, consistent with the long dynamical friction time-scale of L$^{\ast}$ galaxies
\citep{j03}. Hence, apart from the importance of understanding the physical processes
prevailing in transition objects, it is of great significance to find and study  
galaxy groups that show characteristics of a very advanced phase of
evolution such as extremely high density and low velocity dispersion.

In recent work \citep{t03a,t03b}, we have presented a photometric and spectroscopic 
analysis of the ultra-compact galaxy group CG J1720-67.8 \citep{wtk99}. The group, 
characterized by a very high density and the presence of an outstanding tidal 
tail hosting a number of tidal dwarf galaxy (TDG) candidates, reveals properties
indicative of an advanced evolutionary stage. Interpreting its properties in the 
framework of hierarchical clustering theories would suggest that the group is 
close to evolving into to a single merged galaxy, although 
the evolutionary history of this interesting galaxy group is not yet established. 

In this paper we present new observations of CG J1720-67.4 both in the
optical and radio domain and use them in combination with previously obtained 
data to advance our understanding of the evolutionary state of the group. 
In particular, a mosaic of integral field spectra covering virtually the whole extent 
of the group is used to investigate the group kinematics. New 
long-slit spectra of the central/southern part of the tidal tail, obtained at the 
6.5-m Magellan telescope, serves to further investigate the kinematics 
of some tidal dwarf galaxy (TDG) candidates.
The star formation activity of the group is studied using new H$\alpha$ 
and radio-continuum data. H\,{\sc i} observations are used to obtain 
information on the mass of neutral gas in the group, an additional clue to the 
evolutionary state of the system. Finally, we investigate the properties of the
tidal features by means of $BVR$ photometry.

For a description of data obtained at the ESO 3.6 m telescope with EFOSC2
in the $BVR$ bands and their reduction and photometric calibration, we refer the
reader to \citet{t03a}. 
Our observations and analysis of integral field data, long-slit spectra and radio 
data are described in Sections 2, 3, and 4, respectively, where results from the 
individual observations are also reported. Our observational results are used to 
discuss the star formation activity (Section 5), the structure of
the diffuse light and tidal features (Section 6) and the group's kinematics 
and dynamics in the framework of its interaction history (Section 6). 
Our main results are summarised in Section 7. 

\section{INTEGRAL FIELD SPECTROSCOPY}

\subsection{Observations and data reduction}

\begin{figure} 
\includegraphics[width=\columnwidth]{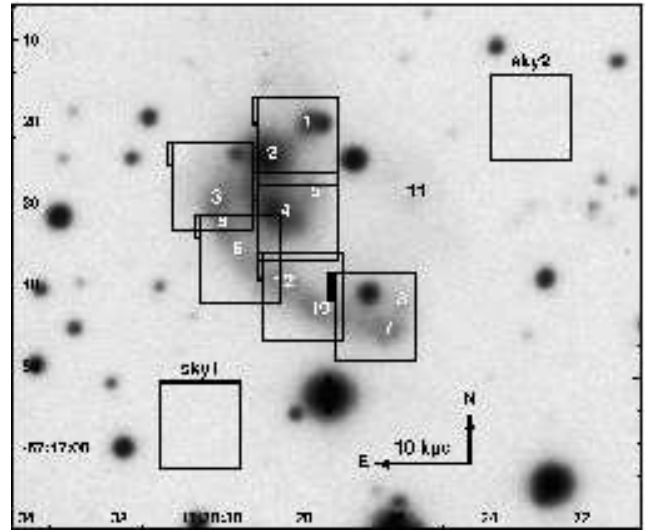}
\caption{Positions of the SPIRAL microlenses array marked on to the ESO 3.6-m telescope 
R-band image of CG J1720-67.8. Objects are labelled according to \citet{t03a,t03b}.
Right ascension and declination (J2000.0) are marked on the axes.}\label{array_pos}
\end{figure}

CG J1720-67.8 was observed in the wavelength range $\lambda$ 6500 - 7120 
\AA\, centered on the redshifted H$\alpha$ line and including 
[N\, {\sc ii}]$\lambda\lambda$ 6548,6583 \AA, and 
[S\,{\sc ii}]$\lambda\lambda$ 6716,6731 \AA\ 
emission lines. A mosaic of fields was made by combining observations carried out in 2002 June and 2003 June 
at the Anglo-Australian Telescope equipped with the integral field unit SPIRAL 
(Segmented Pupil/Image Reformatting Array Lens) and a 600 lines mm$^{-1}$ grating. 
The instrument was used in nod-and-shuffle mode\footnote{See the SPIRAL data reduction manual 
written by J. Bailey, available at 
http://www.aao.gov.au/local/www/spiral/reduction/spiral\_reduction.html} to obtain optimal subtraction 
of the sky-background from the target spectra, especially important for the faintest 
components such as the clumps along the tidal tail.
The useful part of the array encompassed 14$\times$15 micro-lenses\footnote{Each array 
had 4 or 5 additional micro-lenses, which collected useful signal from the target, but
from which the sky-background could not be subtracted.}, giving a field 
of view of 9.8$\times$10.5 arcsec with a spatial scale of 0.7 arcsec per micro-lens
per pixel. Spectra were imaged on to the 2048$\times$4096 13.5 $\mu$m-size pixel EEV 
chip, used in 2 by 2 binning mode, giving a dispersion of 0.6 \AA\ pixel$^{-1}$.
The spectral resolution, evaluated from the FWHM of comparison lines was $\sim$ 80 km s$^{-1}$.
The seeing varied between 1.0 and 
2.2 arcsec during the observations. Nearly the whole extent of CG J1720-67.8 was 
covered with 6 pointings of the array (Fig.~\ref{array_pos}), allowing for sufficient 
overlapping between 
adjacent pointings in order to ensure the reconstruction of the mosaic and the 
consistency of flux calibration. Two series of exposures, alternating sky and target, 
were taken in each pointing for a total exposure time of 30 min on target and
an equal amount on sky. In one of the array pointings,
along the tidal tail of the group, an additional series of exposures was taken in 
order to improve the signal-to-noise ratio.
Observations of the spectrophotometric standard stars LTT4364 and LTT9239 were 
made for flux calibration purposes.

\begin{figure*} 
\includegraphics[width=\textwidth]{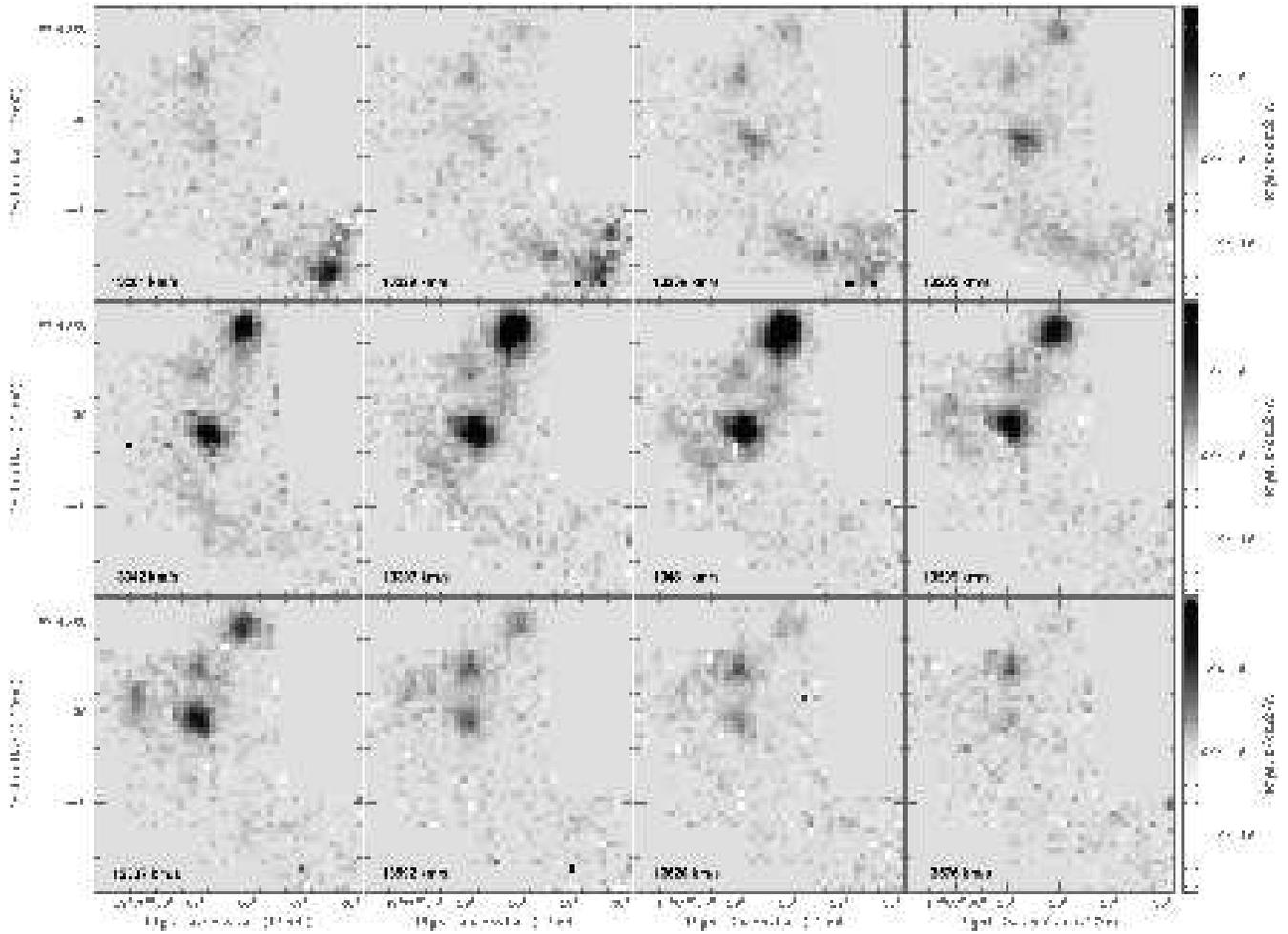}
\caption{First moment monochromatic maps of CG J1720-67.8 at differing radial velocities (referred to
the H$\alpha$ emission-line).}\label{moments}
\end{figure*}

The basic reduction steps, involving the fibre-tracing using flat-field spectra, 
the extraction of the spectra, dark correction, wavelength linearization,
and throughput calibration of the fibres, were done using the 2dfdr software, 
provided by the Anglo-Australian Observatory. Repeated exposures were combined using 
{\sc iraf}\footnote{{\sc iraf} is distributed by the National Optical Astronomy 
Observatory, which is operated by the Association of Universities for Research in 
Astronomy, Inc., under cooperative agreement with the National Science Foundation} 
after correction for atmospheric extinction, flux calibration, and the application 
of a correction for atmospheric telluric absorption bands, particularly affecting 
the [N\,{\sc ii}] $\lambda$ 6583 emission-line. A correction for Galactic extinction
(A$_V$ = 0.29 mag) was also applied.
The exact position of the micro-lens arrays on the target (see Fig.~\ref{array_pos}) 
was verified by reconstructing a continuum image for each array position and measuring 
the position of stars and/or galaxy nuclei falling within the field of view. Spectra 
in the overlapping regions of the arrays were checked for flux calibration consistency. 
Flux calibration of all arrays were found to be mutually consistent and 
an average of the spectra from overlapping fibres was used for the mosaic reconstruction.
An {\sc idl} procedure was written to combine the 6 arrays into a final mosaic and 
{\sc iraf} scripts  for handling of integral field spectra developed and kindly 
provided by M. Radovich and S. Ciroi were modified to reconstruct images of the whole 
mosaic at the desired wavelengths. The calibrated and reconstructed images 
at each wavelength 
step were combined with the {\sc miriad} package into 
a data cube to facilitate the identification of group substructures at different radial 
velocities (see Fig.~\ref{moments}).

For an easier identification of the structures, we projected our reconstructed
maps on to a finer pixel grid (each pixel was projected on to 4$\times$4
pixels) and smoothed the resulting maps. Such maps were used only for
visualization purposes, while measurements were performed on the original data.

A portion of Galaxy 4 including the nucleus was observed in the blue wavelength 
range ($\lambda$ 4850 - 5470 \AA), although under worse seeing conditions 
(the seeing varied from 2.3 to 2.7 arcsec during the exposures). 
The total exposure time on target was 1800 s. Unfortunately, mapping of the rest 
of the group in this spectral range was prevented by unfavourable weather conditions. 
The reduction steps described above were applied to the blue range as well. The flux 
calibration was based on the immediately preceding exposure of the spectrophotometric 
standard star LTT9239. However the sky conditions during the observations were not
photometric. Therefore, a check of the flux calibration was necessary.
Since the position of the array was slightly offset from the 
corresponding position in the red wavelength range, the reconstructed maps in the 
H$\beta$ line and the adjacent continuum were compared to the H$\alpha$ line and 
continuum maps to match the individual spectra in the two ranges. 
A comparison of the continuum level in the two wavelength ranges showed a considerable
discrepancy. Since fluxes in the red part of the spectrum are fully consistent with
fluxes of previously obtained long-slit spectra, the discrepancy was attributed to 
non-photometric sky conditions during the observations in the blue range.
A correction was applied to the blue spectra to match the average continuum level in 
the red part.
The spectra so rescaled have H$\beta$ fluxes consistent with those measured from long-slit
spectra.

\subsection{Measurements and data analysis}
\subsubsection{Intensity and velocity maps}

\begin{figure} 
\includegraphics{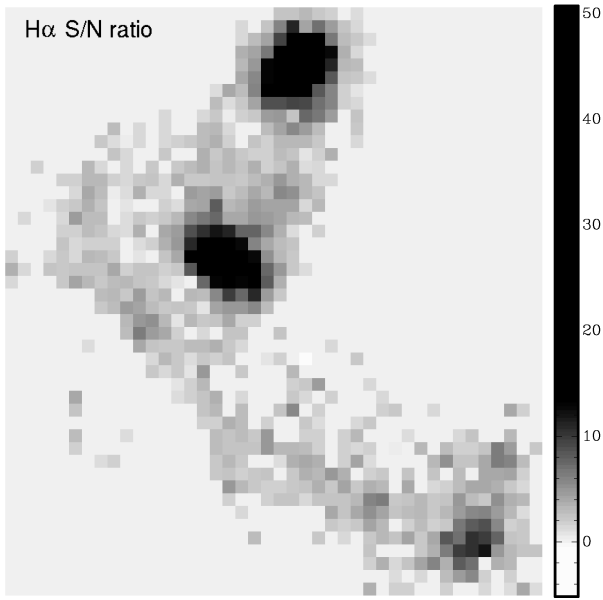}
\caption{Map of the S/N ratio of the H$\alpha$ emission line as measured on the integral-field
spectra. S/N ranges from 1.0 (lightest grey-level) to 50.0 (darkest grey-level).}\label{sn_ratio}
\end{figure}

Emission-line positions and fluxes were determined through Gaussian fitting of the 
lines in the individual spectra. While [N\,{\sc ii}]$\lambda\lambda$ 6548,6583 and 
[S\,{\sc ii}]$\lambda\lambda$ 6716,6731 were detected only in the spectra with best 
S/N ratio, H$\alpha$ was detected in most of the spectra.
A map of the S/N ratio in the H$\alpha$ line is shown in Fig.~\ref{sn_ratio}.
For some portions of the group -- namely the region occupied by TDG3+9, Galaxy 4, and the southern 
part of Galaxy 2 -- it was possible to obtain a (very incomplete) map 
of the emission-line ratio [N\,{\sc ii}]/H$\alpha$. In some cases we found  
[N\,{\sc ii}]/H$\alpha$ slightly higher than expected for H\,{\sc ii} regions, in the
range 0.6 - 1.0, which might indicate the presence of shocked gas. 
The highest values of the emission-line ratio are found in a few positions across TDG3+9,
while moderately high values are found mainly in the outer parts of
Galaxy 4, at the departure of the small tail on its western side and toward Galaxy 2.

\begin{figure} 
\includegraphics{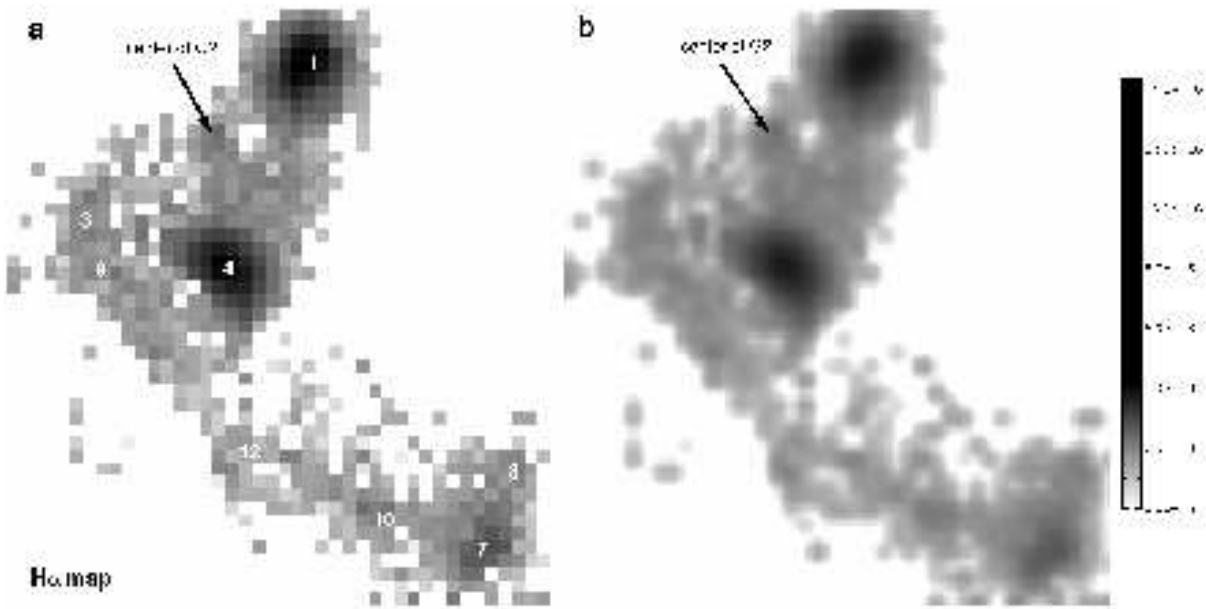}
\caption{(a) H$\alpha$ map of CG J1720-67.8, reconstructed from flux measurements of the emission-line
in the integral field spectra. Object are labelled according to \citet{t03a}. (b) Same map projected 
on to a finer pixel grid.}\label{Ha_mosaic}
\end{figure}

\begin{figure} 
\includegraphics{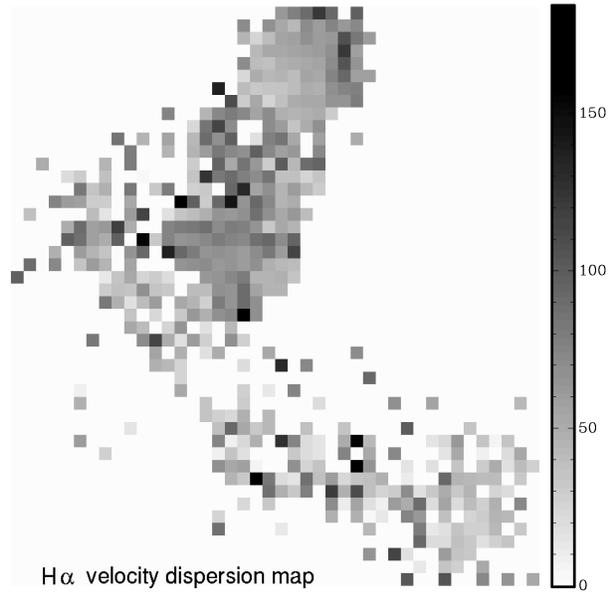}
\caption{Map of the gaseous velocity dispersion calculated from the intrinsic FWHM of the H$\alpha$ 
emission-line. Where measurable, values range from $\sim$ 6 (lightest grey-level) to $\sim$ 180 
(darkest grey-level) km s$^{-1}$.}\label{mosaic_sigma}
\end{figure}

H$\alpha$ measurements were used to obtain a map of the group in the emission-line 
(Fig.~~\ref{Ha_mosaic}) and a radial velocity field of the ionized gas. The intrinsic width of the line 
varied across the objects from instrumental values up to $\approx$ 10 \AA.
A map of the velocity dispersion $\sigma_{\mathrm {gas}}$ of the gas 
(Fig.~\ref{mosaic_sigma}) was derived from the H$\alpha$ FWHM, after 
correction for the instrumental FWHM, as measured on the comparison spectrum in the 
relevant fibres, in order to take into account linewidth variations across the array. 
In some points $\sigma_{\mathrm {gas}}$ reaches values $\approx$ 180 km s$^{-1}$.
Interestingly, the highest values of velocity dispersion are found in the 
central/southern part of Galaxy 2 and in the northern part of Galaxy 4, where the 
two galaxies partly overlap. The broadening of the emission-line in this region
might indicate the presence of an unresolved second component at slightly different 
radial velocity, originating in Galaxy 2. Additional regions showing particularly high 
gaseous velocity dispersion are the western side of Galaxy 1, the central part of the 
TDG candidate 3+9, and a region south of the nucleus of Galaxy 4, where the presence 
of a clump was evident from previous optical photometry \citep[fig.~5 in ][]{t03a}.

\begin{figure} 
\includegraphics[width=\textwidth]{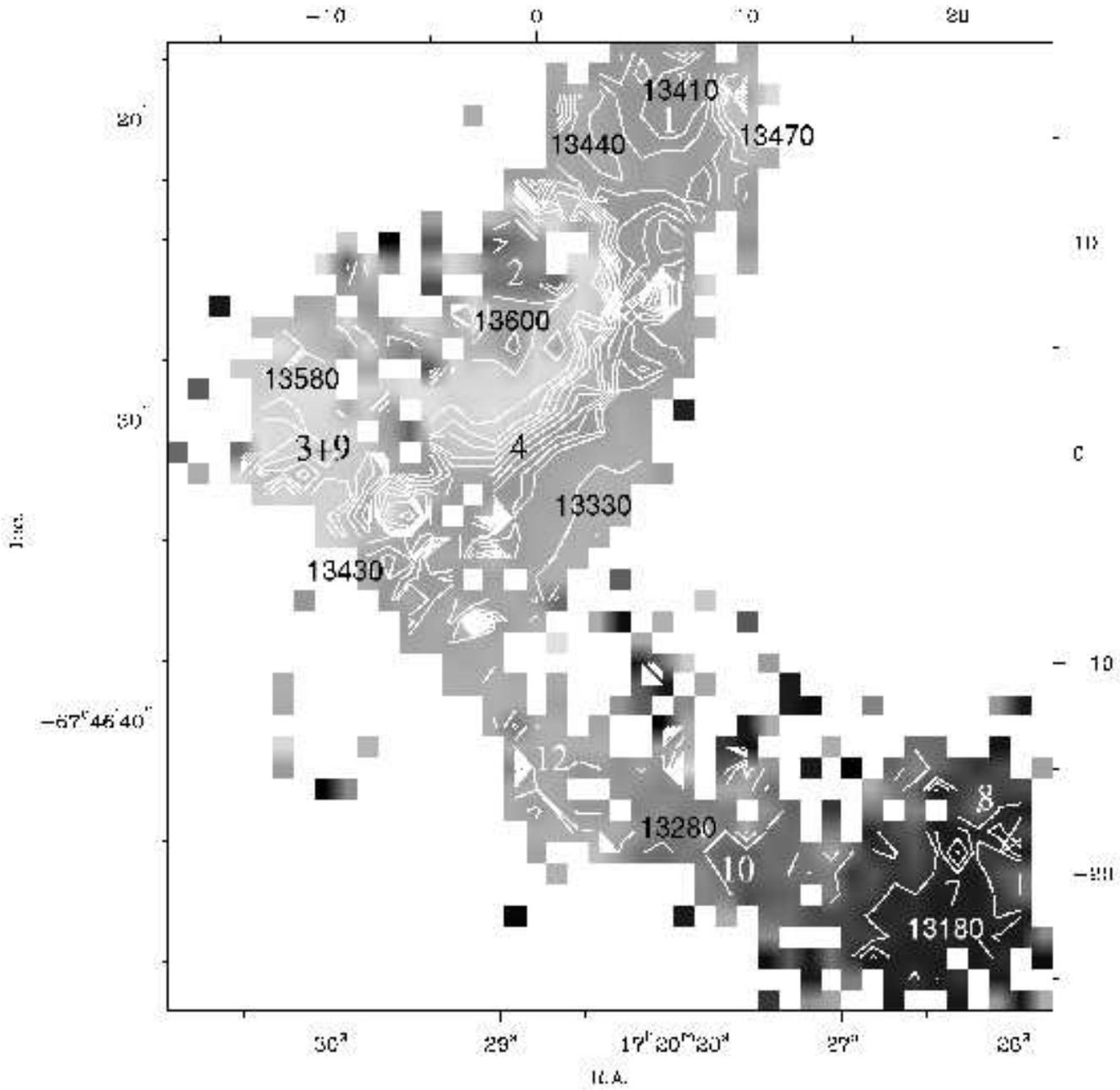}
\caption{H$\alpha$ velocity field of CG J1720-67.8. Velocity contours are overplotted and some reference
values in km s$^{-1}$ are marked. Galaxy positions are indicated. A clear velocity gradient with 
velocities increasing from south-west to north-east is visible. The kinematic axis of Galaxy 4 is
clearly visible at a position angle (P.A.) of $\approx$ 40\degr.}\label{mosaic_hevel}
\end{figure}

The positions of 6 night-sky emission-lines were measured in every spectrum of the six 
different arrays and compared with predicted wavelengths. No systematic dependences of 
the error on the position within an individual array were detected.
The median error in radial velocity ranges from 2.3 to 5.1 km s$^{-1}$ depending on 
the array considered. Errors in radial velocities are known to depend on the S/N ratio of the lines 
used for the measurements. We verified that measured position of low-S/N night-sky lines 
yielded to errors $\la$ 20 km s$^{-1}$. We adopt this value as an estimate of our error 
in radial velocity for regions where the S/N ratio of the H$\alpha$ emission-line was 
particularly low (see Fig.~\ref{sn_ratio}). A mosaic of the group's velocity field 
(Fig.~\ref{mosaic_hevel}) was reconstructed after application of the relevant heliocentric 
correction for each array.

\begin{figure} 
\includegraphics{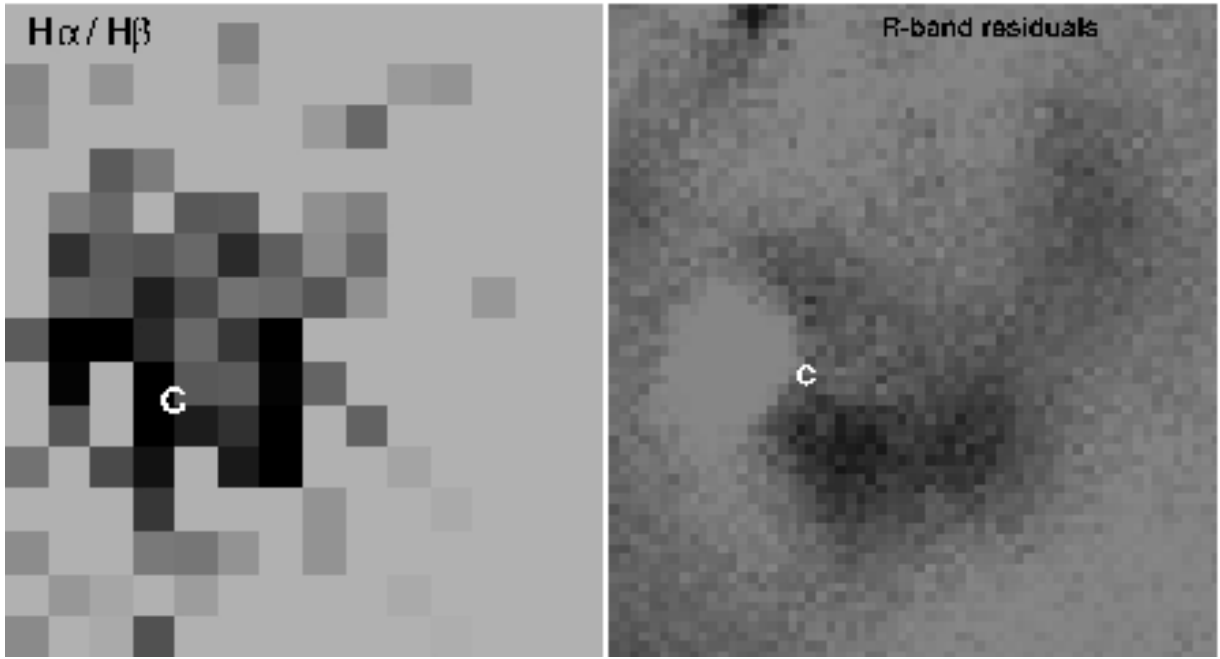}
\caption{\emph{Left:} Extinction map of Galaxy 4. The center of the galaxy is marked with `C'. 
North is up, east to the left. Higher values of extinction are represented with darker grey-levels.
\emph{Right:} The same region on the R-band image, in inverted grey-scale, after subtraction of a 
bidimensional model of the light distribution of Galaxy 4. The residuals of light north-east and
south-west of the nucleus roughly coincide with the positions of higher extinction.}\label{Ha_Hbratio}
\end{figure}

For the portion of Galaxy 4 observed both in the blue and red wavelength range, 
we were able to obtain a map of the internal extinction in terms of H$\alpha$/H$\beta$ 
ratio (Fig.~\ref{Ha_Hbratio}). The observed ratio has values between $\sim$ 3.5 and 8.5
around the center of the galaxy and is particularly high in regions north-east and south-west 
of the center, where it varies between $\sim$ 7.0 and 10.5. Assuming a case B photoionized 
nebula and a \citet{ccm} extinction law, this translates into a color excess E($B-V$) in the 
range $\sim$ 0.2 to 1.3 mag across the galaxy. 
Extinction estimates become uncertain in the outer parts of the galaxy, where the S/N ratio 
of the emission lines (in particular of H$\beta$) decreases.
These results confirm our earlier interpretation of the galaxy
broad-band color indices \citep{t03a}. Interestingly, as visible in Fig.~\ref{Ha_Hbratio},
the position of highly extinguished
regions is roughly coincident with the residual structures observed in broad-band images after 
subtraction of the best-fitting bi-dimensional galaxy model and interpreted in our earlier 
work as regions of enhanced star formation.

\subsubsection{Analysis of the velocity field}

As visible in Fig.\ref{mosaic_hevel}, the entire group appears dominated by the kinematics of
Galaxy 4, although, at first glance, the kinematic center seems to be displaced from the
photometric center of the galaxy. Galaxy 1 shows the same systemic velocity as Galaxy 4,
while the systemic velocity of Galaxy 2 is roughly 200 km s$^{-1}$ higher.
To investigate the kinematics of the three main galaxies, we attempted a fitting of iso-velocity 
tilted rings to the bi-dimensional velocity field by applying the {\sc gipsy} task {\sc rotcur}, 
following the procedure described by \citet{beg89}.

The velocity field of Galaxy 1 shows little gradient and appears significantly distorted.
The orientation of the kinematic axis is not clear, although the small velocity gradient
observed might indicate a  P.A. $\approx$ 160\degr. By fixing the centre position and disc inclination
to photometrically determined values, the fitting procedure gave a rather stable value of the
systemic velocity V$_{\mathrm{sys}}$ = 13422 $\pm$ 2 km s$^{-1}$. 
However the attempting to fit rotational velocities at differing radii produced randomly varying 
values and hence was not successful. No velocity curve could be derived for this galaxy.

Galaxy 2 shows some velocity gradient, but the ionized gas was detected in too small a region 
to allow a fitting of the velocity field and the extraction of a velocity curve.

\begin{figure}
\includegraphics[width=8cm]{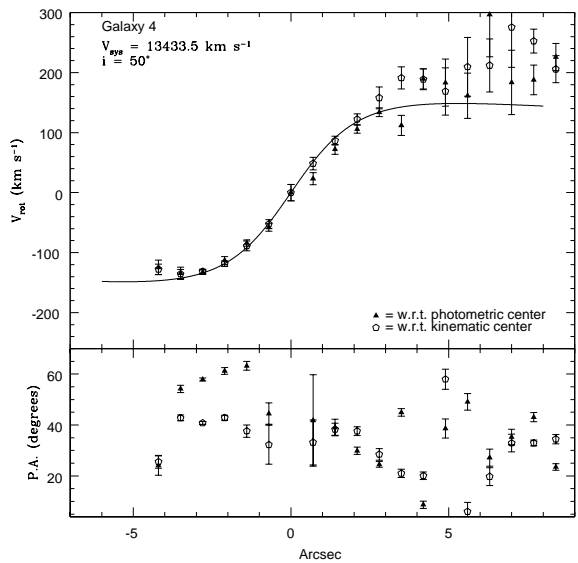}
\caption{Rotation curve of Galaxy 4 (upper panel) derived by fitting concentric tilted rings to the velocity
field using as centre position the photometric centre (triangles) and the kinematic centre
(penthagons). The solid line represents the best fitting curve to the de-projected circular 
velocities. The lower panel shows the radial trend of the kinematic major axis P.A. for both
choices of centre position.}\label{g4rotcurve}
\end{figure}

The procedure was successfully applied to Galaxy 4, allowing us to obtain its rotation curve.
The fitting was performed in several steps. To limit the number of free
parameters we adopted as inclination value for the galaxy $i$ = 50\degr\ based on our photometric
analysis. 
The systemic velocity was determined as average of the V$_{\mathrm{sys}}$
of individual concentric tilted rings with fixed inclination, centred on the
photometric centre of the galaxy. This value, V$_{\mathrm{sys}}$ = 13433.5 $\pm$ 13.7 km s$^{-1}$,
was adopted for all rings during the subsequent iterations of the fitting.
In the second step we allowed the position of the centre to vary during the fitting, to determine the
kinematic centre. The fitted centre position changes for different rings and tends to move away
from the photometric centre in a north-west direction along the kinematic minor axis,
thus confirming the impression that the kinematic centre is considerably offset with respect to the
photometric centre of Galaxy 4. We adopted as the kinematic centre the fitted position for the
iso-velocity ring with radius 2.1 arcsec. The offset between the photometric centre and the adopted
kinematic centre is $\sim$ 5.9 arcsec. Finally we fitted separately the approaching and receding side 
of the galaxies maintaining V$_{\mathrm{sys}}$, $i$, and the centre position fixed. The fitting
procedure was performed once with respect to the photometric centre and once with respect to the kinematic
centre,  only including points within $\pm$25\degr of the kinematic major axis. This choice
made it possible to include a sufficient number of points and, at the same time, minimise the 
contamination by adjacent galaxies. The resulting (de-projected) rotation curve is shown in Fig.~\ref{g4rotcurve} for both 
choices of centre position. The fitting of the approaching side succeeded only up to a radius 
$r$ $\sim$ 4.2 arcsec, beyond which
not enough points are available and the contamination by the nearby tidal tail becomes strong.
The fitting of the receding side could be extended to a radius of 8.4 arcsec, however for $r$ $\geq$ 5 arcsec
the curve becomes irregular because of contamination by Galaxy 2, which lies north of Galaxy 4 
and partly overlaps to it. The curves obtained with the two different centres are very similar.
However the one referred to the kinematic centre is better behaved. The rotation curve is symmetric 
out to a radius of 2 to 3 arcsec, beyond which the two halves tend to diverge.

In order to estimate the mass of the galaxy we attempted to fit the rotation curve (for radii $r$ $<$ 5
arcsec) assuming pure circular motion of the gas. Following \citet{b91} we used as a fitting
law for the circular velocity V$_{\mathrm{c}}$
\begin{equation}
\mathrm{V_c} = A r (r^2 + c_0^2)^{-p/2},
\end{equation}
where the parameters $A$, $c_0$, and $p$ were allowed to vary.
The fitting was performed with a least-squares Levenberg-Marquardt algorithm and gave as a result
$A$ = 265 km s$^{-1}$, $c_0$ = 2.7 arcsec, and $p$ = 1.26 with an rms = 28 km s$^{-1}$ and a reduced $\chi^2$ =
1.9. The fitted rotational velocity at $r$ = 5 arcsec, where the curve tends to flatten, is
V$_{\mathrm{c}}$(5\arcsec) = 148 km s$^{-1}$. The mass within this radius can be estimated through the
expression
\begin{equation}
\mathrm{M = G}^{-1} r \mathrm{V_c}^2(r),
\end{equation}
which yields M(5\arcsec) = (2.2 $\pm$ 0.8)$\times$10$^{10}$ M$_{\odot}$. 
This estimate of the mass of Galaxy 4 is in good agreement with the value obtained through a comparison of the
observations with spectrophotometric evolutionary synthesis models \citep{tf03}.

Since the length of the tail, measured along its spine on the optical 
images, is $\sim$ 28.8 kpc, a rough estimate of its age is given by t$_{\mathrm{tail}}$ 
$\approx$ 28.8 kpc/148 km s$^{-1}$ $\approx$ 200 Myr, under the assumption that the tidal 
tail has been launched by Galaxy 4. We note that, since V$_{\mathrm{c}}$(5\arcsec) is only a 
lower limit to the maximum rotational velocity, the above estimate is an upper 
limit to the age of the tail. The age of the tail, therefore, suggests a relatively recent
interaction.
For comparison, the estimated ages of the tidal tails 
in a sample of merging systems along the Toomre sequence \citep{too77} range between 400 
and 770 Myr, the highest value being found for NGC 7252, the system located at the end of 
the sequence \citep{kn03,hi94}. 
Interestingly, the age of the interaction-induced burst of star formation of 
Galaxy 4, as estimated from comparison of the observations with spectrophotometric 
evolutionary synthesis models, is $\sim$ 40 -- 180 Myr \citep{tf03}, the oldest age offering
a better match to the observations, consistent with the above estimate of t$_{\mathrm{tail}}$.

\subsubsection{Total spectra of individual objects}

\begin{figure} 
\includegraphics{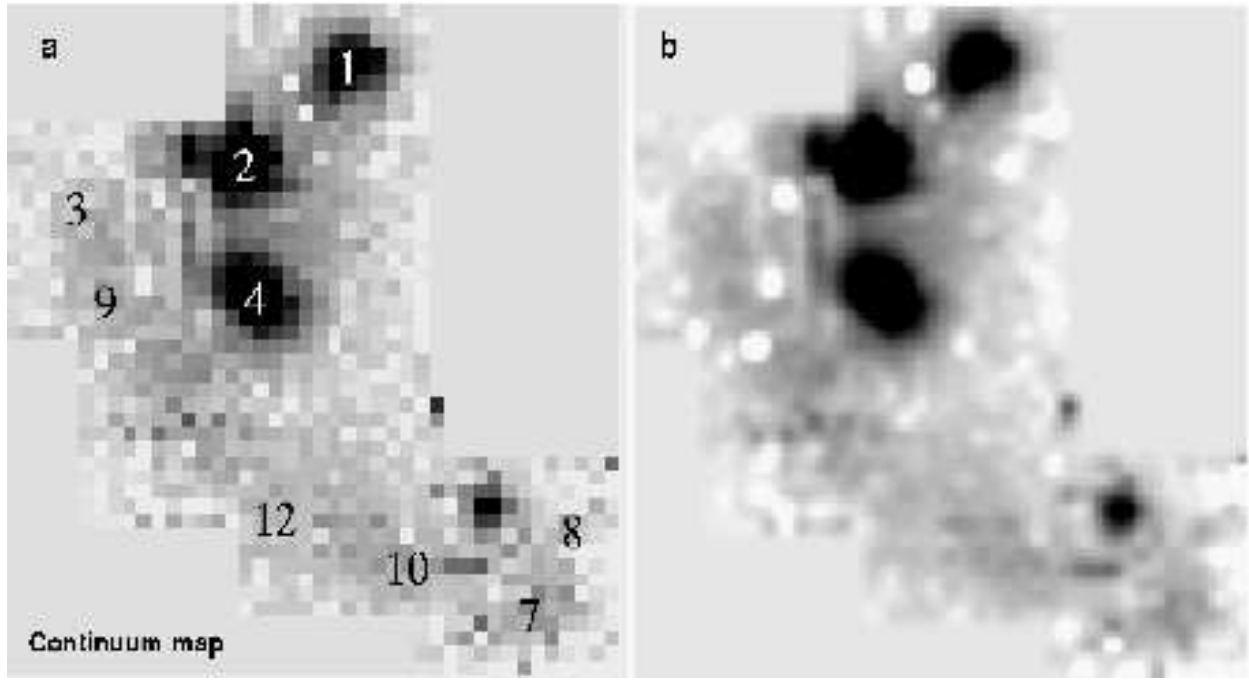}
\caption{(a) Continuum map of CG J1720-67.8 obtained from integral field spectra by integrating
the flux in the range $\lambda$ 6500 -- 6800 \AA\ and reconstructing the mosaic of the group. 
(b) The same map projected on to a finer pixel grid and adequately smoothed, for better visibility 
of the structures.}\label{cont_map}
\end{figure}

Fibre-spectra belonging to each individual object were identified on the continuum map 
(Fig.~\ref{cont_map}), and added together to obtain the relevant total spectra. 
These are more suitable 
than long-slit spectra to derive average physical properties of the ionized gas in each 
object and to estimate the star-formation activity of each galaxy.
Relevant measurements are listed in Table~\ref{emlines_tot}. 
The H$\alpha$ fluxes in Table~\ref{emlines_tot} are somewhat different from those derived 
in Section 5.1 through polygonal aperture photometry.
These 15 - 20 per cent discrepancies are due to difficulties in the aperture definition
in case of partly overlapping sources. 
The diagnostic emission-line ratios [N\,{\sc ii}]/H$\alpha$ and [S\,{\sc ii}]/H$\alpha$
agree, within the errors, with those plotted on fig.~10 of \citet{t03a}.
In addition, a comparison of our new measurements of H$\alpha$ fluxes with previous ones 
shows that only in the case of Galaxy 4 slit-spectra significantly underestimated 
(by about 67 per cent)
the total flux, while for the other galaxies the bulk of the flux fell within the slit.
Therefore star formation properties of the individual objects remain unaltered with respect
to our earlier claims, with the exception of Galaxy 4, for which a total star formation
rate twice as high as the previously determined one is suggested 
(i.e. $\approx$ 3 M$_{\odot}$ yr$^{-1}$, after correction for internal extinction).

An integrated spectrum of Galaxy 4 was obtained in each wavelength range by adding up 
all spectra covering the same portion of galaxy in both ranges.
We measured an average internal extinction  E($B-V$) $\sim$ 0.6 mag, consistent with that
obtained from long-slit spectra \citep[][table 2]{t03a}.
The presence of dust patches causing extinction in excess of the average E($B-V$) 
determined for the galaxy (Section 2.2.1) suggests that the overall H$\alpha$ luminosity and 
star formation
rate of the galaxy might be somewhat underestimated even after the application of an average 
internal extinction correction.

\section{LONG-SLIT SPECTROSCOPY}

\subsection{Observations and data reduction}

A new long-slit spectrum, across the clumps in the southern part of the tidal tail of 
CG1720-67.8 (position angle 73\fdg5), was obtained in 2002 May at the Magellan 6.5-m 
telescope (Las Campanas Observatory). This spectrum complements the data previously 
obtained in other regions of the group, and already described in \citet{t03a,t03b}. 
The 1800 s exposure was taken through a 72.0$\times$1.25 arcsec slit with the 
Boller and Chivens spectrograph and a 600 lines mm$^{-1}$ grating, and imaged on to a 
2048$\times$515 13.5-$\mu$m pixel CCD. This configuration gave a dispersion of 1.557 
\AA\ pixel$^{-1}$ in the wavelength range 3780 - 6970 \AA, a spectral resolution of 
$\sim$ 6.8 \AA\ (FWHM), and a spatial scale of 0.25 arcsec pixel$^{-1}$. The seeing 
at the time of the observations was $\sim$ 1.2 arcsec.

The usual reduction steps: bias subtraction, flat-fielding, and wavelength calibration, 
were carried out on the bi-dimensional spectrum with standard {\sc iraf} packages, while 
the {\sc iraf} script 
{\sc l.a.cosmic}\footnote{Available at http://www.astro.yale.edu/dokkum/lacosmic/}
\citep{vdokk01} was used for cosmic ray rejection.
An average sensitivity curve obtained from three exposures of the standard star CD32d9927,
taken during the same night, was used for the absolute flux calibration.
The spectrum shows the usual Balmer hydrogen emission-lines, [O\,{\sc ii}] $\lambda$ 3727,
[O\,{\sc iii}] $\lambda$ 4959,5007, He\,{\sc i} $\lambda$ 5876, [O\,{\sc i}] $\lambda$ 6300,
and [N\,{\sc ii}] $\lambda$ 6548,6583.

\subsection{Measurements and data analysis}

\begin{figure} 
\includegraphics{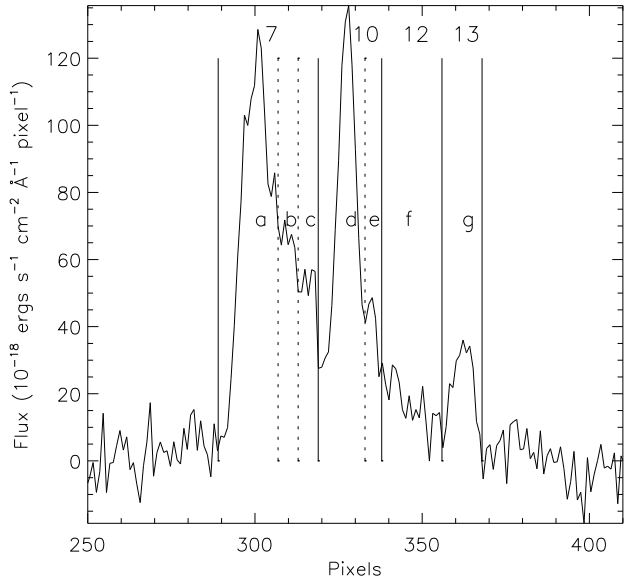}
\caption{Continuum-subtracted H$\alpha$ emission-line profile along the slit. Numbering is consistent
with \citet{t03a,t03b}. An additional region unidentified in previous works is labelled
`13'. Small case letters and dashed lines indicate subregions identified on the present profile.}\label{slit_profile}
\end{figure}

Individual emission regions were identified on the continuum-subtracted H$\alpha$ 
emission-line profile along the slit (Fig.~\ref{slit_profile}). 
Some of them are portions of the TDG candidates found in the central and southern parts of 
the tidal tail \citep{t03b}. We labelled these regions 
according to our previous papers (see also Fig.~\ref{array_pos})
although we must stress that the spectra previously obtained sampled different parts
of these clumps, hence emission-line fluxes (and ratios) are not expected to be exactly the same.
An additional, fainter emitting region, labelled `13' is visible in the spectrum 
and appears to correspond to a faint clump lying to the east of object 12.
One-dimensional spectra of the individual regions were extracted and corrected for 
Galactic extinction (A$_V$ = 0.29 mag). 
Once the spatial coverage of the slit has been taken in to account,the fluxes of
the detected emission-lines are in reasonable agreement,
as an order of magnitude, with those measured on integral field data, 
therefore we do not quote them.

\begin{figure} 
\includegraphics{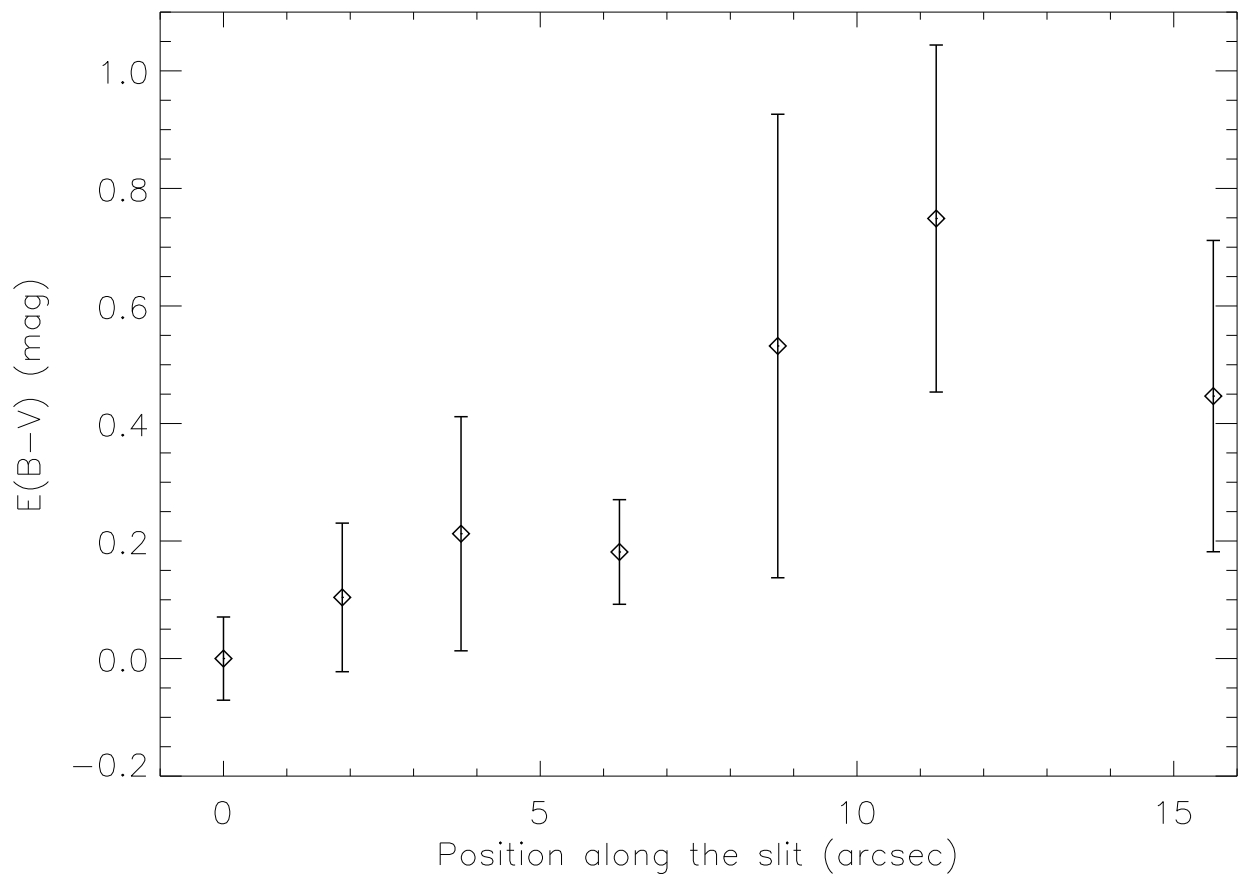}
\caption{Trend of the internal extnction, expressed in terms of E($B-V$), along the slit on
the southern part of the tidal tail.}\label{ebv-slit}
\end{figure}

The emission-line ratios obtained with the new spectrum confirm the physical conditions 
and metal abundances (ionization parameter $\log$U $\approx$ $-$3, Z $\sim$ 0.2 - 0.3 Z$_{\odot}$) 
already suggested for the TDG candidates, with a tendency for TDG 7+8 toward lower metallicities. 
An interesting outcome of the new observations concerns the distribution of dust across
the star-forming clumps. In fact considerable variations of E($B-V$) along the slit reveal
an unhomogeneous distribution of dust in the southern part of the tidal tail, as shown in
Fig.~\ref{ebv-slit}. In particular, we observe a variation of E($B-V$) from $\sim$ 0.0 to 0.2 mag
in the region occupied by TDG 7. The Balmer decrement on a spectrum across the same object but at 
a different position angle gave E($B-V$) $\sim$ 0.4 mag \citep{t03a}. 

\begin{figure} 
\includegraphics{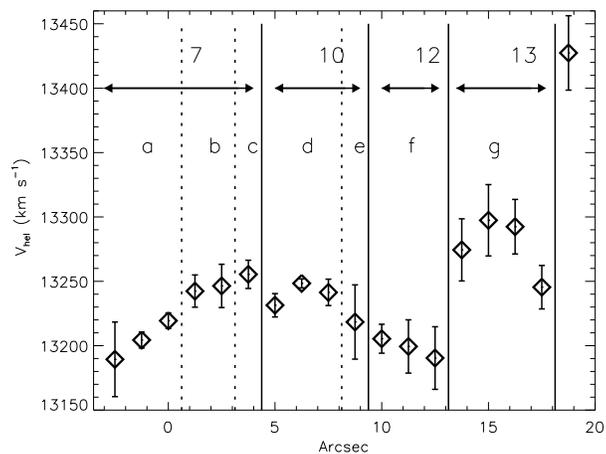}
\caption{Velocity gradient measured from the H$\alpha$ emission-line along the slit. Solid lines separate
points belonging to objects 7, 10, 12, and 13. Small case letters and dashed lines indicate sub-regions
isolated on the H$\alpha$ profile of Fig.~\ref{slit_profile}.}\label{velcurve}
\end{figure}

The brightest emission lines were used to obtain 
independent radial velocity measurements along the slit 
by binning the spectrum along the spatial direction according to the seeing. 
The resulting radial velocity trend is
shown in Fig.~\ref{velcurve} and will be further discussed in Section 7.2.
Errors in radial velocity are estimated from the average shift of night-sky lines with 
respect to the expected wavelengths along the slit.
Sub-regions marked in Fig.~\ref{velcurve} do not necessarily indicate
distinct objects; in fact there is no clear kinematic distinction between them.
However, it is interesting to note that there are kinematic discontinuities
between the different clumps cross-identified with TDG candidates. Within each
clump we observe regular velocity gradients, sometimes reversed with
respect to the general velocity gradient observed along the tidal tail, visible
in Fig.~\ref{velcurve} and, more clearly, in the velocity field of
Fig.~\ref{mosaic_hevel}. Within the clumps we measured velocity gradients of
order of $\approx$ 20 km s$^{-1}$ kpc$^{-1}$, similar to gradients observed by
some authors in other TDG candidates \citep[e.g., ][]{mdo01}.

\section{RADIO OBSERVATIONS}

\subsection{H\,{\sevensize\bf I} observations}

\begin{figure*}
\includegraphics{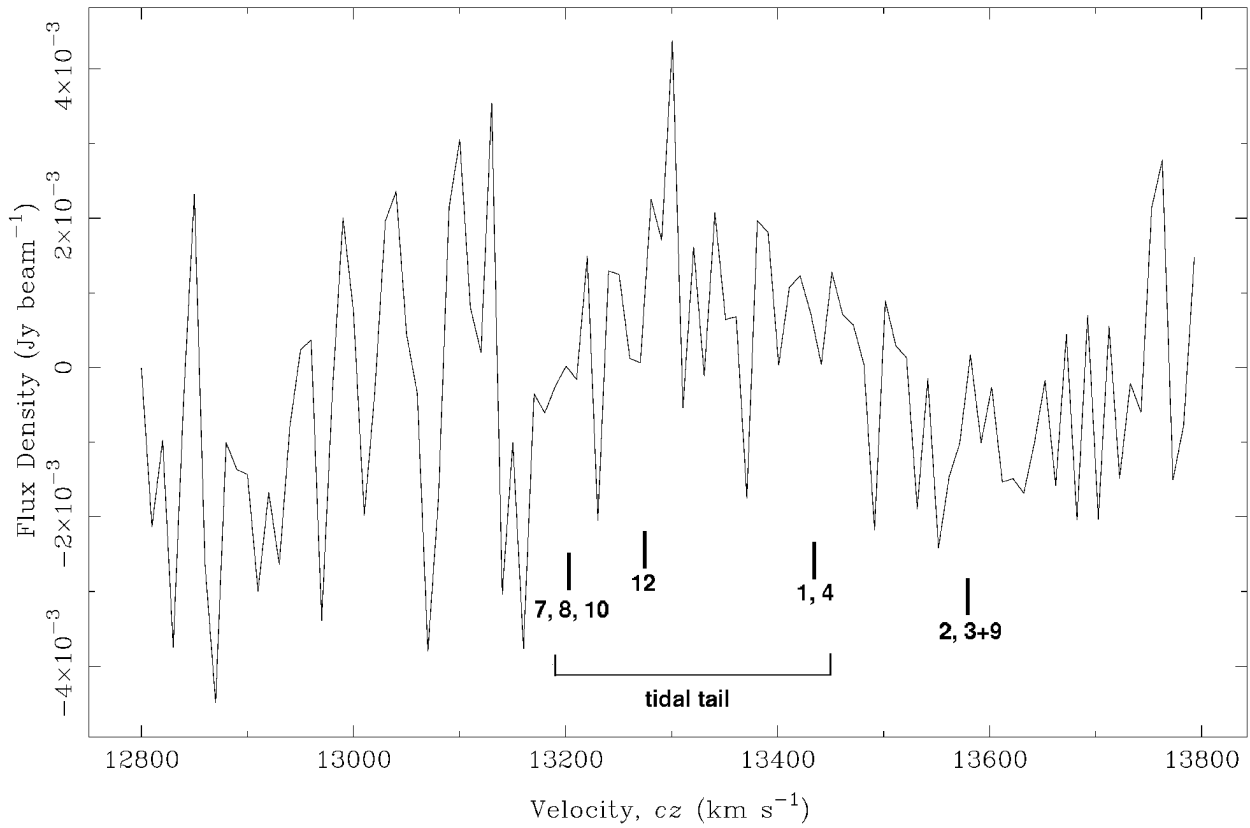}
\caption{ATCA H\,{\sc i} spectrum at the position of CG1720-67.8. Velocity positions of the group members, as
derived from optical observations, are marked.}\label{radio-spec}
\end{figure*}

Observations were made in 2002 January 17 and February 25 with the
750A and 1.5A configurations of the Compact Array\footnote{The Compact
Array is part of the Australia Telescope which is funded by the
Commonwealth of Australia for operation as a National Facility managed
by CSIRO}, respectively. The integration time was $2\times12$ hrs, and
the frequency of the redshifted H\,{\sc i} line was predicted to lie at 1359
MHz. The bandwidth was 8 MHz, and the channel spacing was 15.6 kHz.
Images were made using `natural' weighting, resulting in a beam size
(FWHP) of around 44 $\times$ 39 arcsec. The rms noise was 1.5
mJy beam$^{-1}$ when the data were smoothed to 10 km s$^{-1}$ (close to the
theoretical value) and 0.5 mJy beam$^{-1}$ when smoothed to 200 km s$^{-1}$.
No H\,{\sc i} emission was detected anywhere in the velocity range $cz=12800$
to 13800  km s$^{-1}$ (see Fig.~\ref{radio-spec}). We estimate an 3-$\sigma$ 
upper limit to any emission (assuming a velocity width of 200 km s$^{-1}$) 
of 0.3 Jy  km s$^{-1}$. At a distance of 180 Mpc, this corresponds to 
M$_{\mathrm {HI}}<2.3\times10^9$ M$_{\odot}$. The observational parameters are 
summarised in Table~\ref{radio-obslog}, and the results are summarised in 
Table~\ref{HIobs}.

\subsection{Radio continuum observations}

Radio continuum observations were made at a frequency of 1384 MHz on
2002 January 17 and February 25 with the 750A and 1.5A configurations
of the ATCA (simultaneous with the H\,{\sc i} observations described above).
Additional observations were taken in two sessions on 2003 August 4
and 5 using the 6D configuration. On the first day, two frequencies
centered at 4800 MHz and 8640 MHz were observed simultaneously over 12
hrs. On the second day, we switched between two pairs of frequencies
at 5184 and 5952 MHz and 1384 and 2368 MHz over the observing period
of 11 hrs. In each case, the bandwidth of the radio continuum
observations was 128 MHz, and full polarimetry was obtained. As for
the H\,{\sc i} observations, the secondary (phase and amplitude) calibrator
was PKS B1718-649, and the primary (flux scale) calibrator was PKS
B1934-648.  The observational parameters are summarised in Table~\ref{radio-obslog}.

Images were made using `uniform' weighting at 1384 MHz. But, for
maximum sensitivity, the images were made with `natural' weighting at
other frequencies. The beamsize and RMS noise at each frequency is
listed in Table~\ref{cont-obs}. In each case, CG J1720-67.8 was detected. At 1.4
GHz, the integrated flux density is 4.2 mJy. At 5 GHz, the
integrated flux density is around 1.1 mJy, or slightly higher (the
smaller beam and lower S/N ratio means that some extended emission
will not have been included). This implies a spectral index
($S\propto\nu^{\alpha}$) of $\alpha=-1.0$, consistent with
synchrotron radiation. 

\begin{figure*}
\includegraphics[width=\textwidth]{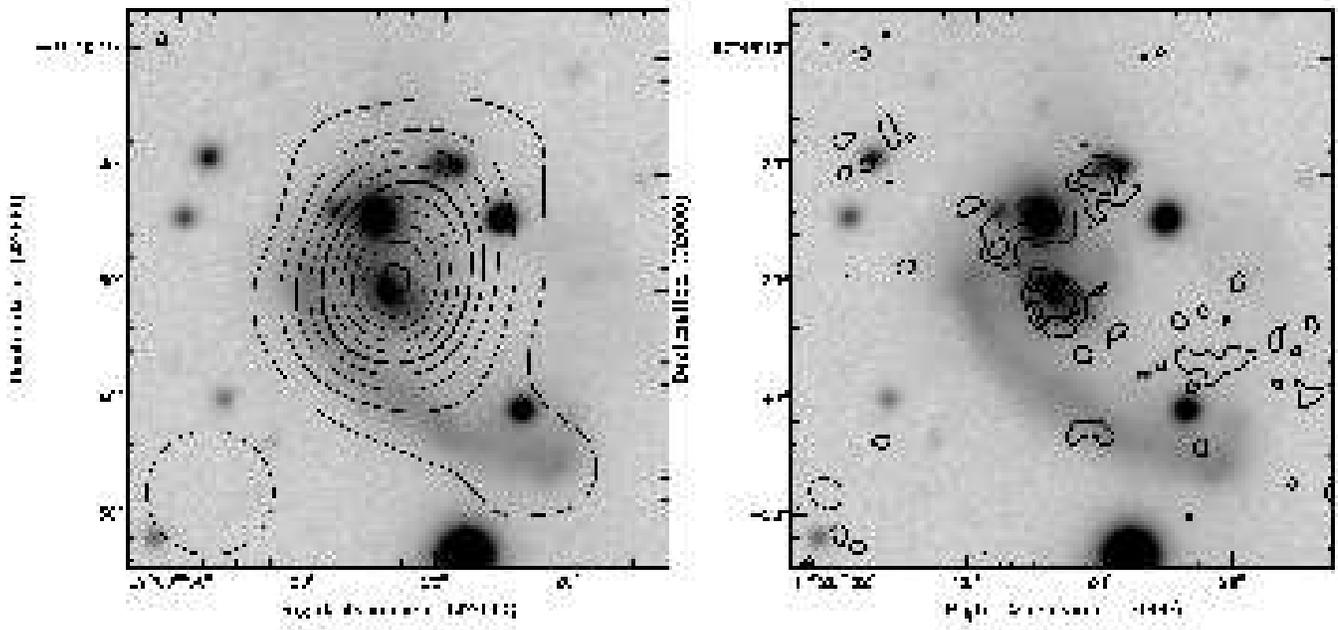}
\caption{ ESO 3.6-m R band image of CG J1720-67.8 overlaid with ATCA 1.4-GHz (a) and 5-GHz (b) 
contours. The ATCA beams are shown on the lower left corners.}\label{radio-contour}
\end{figure*}

As Fig.~\ref{radio-contour} and Table~\ref{cont-obs} show, the emission 
at all frequencies peaks at 17$^h$20$^m$28$^s$.8, $-$67\degr46\arcmin31\arcsec\  
(J2000), which is within 1 arcsec of the optical
centroid of Galaxy 4 \citep{t03a}.  The 1.4-GHz image shows
a significant extension of the radio source to the north towards Galaxy 1, but the outer
contours also reveal a contribution from Galaxy 2, which is seen more
clearly in the 5-GHz image. At 1.4 GHz, there is also a significant
emission (approximately 0.2 mJy) from the faint candidate TDG 7 at the end of
the tidal arm.

\section{STAR FORMATION}

\subsection{Overall star formation properties of CG J1720-67.8}

The total observed H$\alpha$ flux of the group, determined by adding
up the fluxes from each pixel in the H$\alpha$ map
(Fig.\ref{Ha_mosaic}) is F(H$\alpha$) = 9.5 $\times$ 10$^{-14}$ ergs
s$^{-1}$ cm$^{-2}$. At the distance of the group, d = 180 Mpc, this
gives the luminosity L(H$\alpha$) = 3.7 $\times$ 10$^{41}$ ergs
s$^{-1}$. Note that this luminosity is not corrected for internal
extinction, therefore it is to be considered as a lower limit to the
H$\alpha$ luminosity of the group. The value is slightly higher but
consistent with our previous measurements based on long-slit
spectroscopic observations which give a combined H$\alpha$ luminosity of
$\sim$ 3.3 $\times$ 10$^{41}$ ergs s$^{-1}$ for the condensed
components (galaxies, tidal features, tidal dwarf candidates) in CG
J1720-67.8 \citep[][tables 2 and 3]{t03a}.  The internal extinction
values determined from the Balmer-decrement in the long-slit spectra
give an average color-excess $E(B-V)$ = 0.5$\pm$0.2 mag. This, applied
to the total observed flux, yields an extinction-corrected total
luminosity L(H$\alpha$)$_{\mathrm{corr}}$ = 1.2 $\times$ 10$^{42}$
ergs s$^{-1}$, which translates in to a total present-day
star-formation rate \citep{k98} SFR = 10 M$_{\odot}$ yr$^{-1}$ over the
stellar mass range 0.1 -- 100 M$_{\odot}$.

This is somewhat lower than the value of 42 M$_{\odot}$ yr$^{-1}$
derived from the far-infrared flux \citep{t03a}. It is possible that
the H$\alpha$ extinction has been greatly underestimated. On the other
hand \citet{t03a} note that far-infrared measurements in the
vicinity of CG J1720-67.8 are confused by infrared cirrus.  
However, it has also been pointed out by \citet{blr89} that the 
60-to-100 $\mu$m
flux density ratio f$_{60}$/f$_{100}$ can be used as an indicator of
the dust heating source, and therefore the efficacy of far-infrared flux as
an estimator of SFR. The ratio for
CG J1720-67.8 is f$_{60}$/f$_{100}$ = 0.3 $\pm$ 0.1, which is closer to the
range f$_{60}$/f$_{100}$ $<$ 0.3 where dust heating is dominated by the 
diffuse interstellar radiation field of old disc stars, than the
range f$_{60}$/f$_{100}$ $\geq$ 0.5 where
heating is entirely dominated by the UV radiation from young stars, and where
far-infrared flux is a good SFR indicator. This suggests that the 
far-infrared-derived SFR may be an upper limit.

As discussed by many authors \citep[e.g.][]{cr98}, radio flux
densities of galaxies without AGN activity provide yet another
method of estimating star-formation rate, but without the problems of
extinction. At 1.4 GHz, the total flux density of 4.2 mJy translates
in to a star-formation rate of 18 M$_{\odot}$ yr$^{-1}$ using the
relationship established by \citet{c92} and \citet{ha00} with
$Q=5.5$. This is slightly less than double the H$\alpha$-derived SFR
and slightly less than half of the far-infrared-derived SFR. The
discrepancy between the radio and H$\alpha$-derived SFRs appears to
be no more than the scatter observed in other local galaxies \citep{cr98}, 
and the total extinction of $E(B-V) = 0.78$ mag, required to
make the radio and H$\alpha$ SFRs in exact agreement, compares reasonably
with the Balmer-decrement value of $E(B-V)$ = 0.5$\pm$0.2 mag.

Total H$\alpha$ fluxes of the individual group components can 
however be obtained from the final map with better resolution than the
radio data allow. We measured separately the main
tidal tail (including the candidate TDGs), galaxies 1, 2, and 4, and
the secondary short tidal tail west of Galaxy 4 (including object 5)
by means of polygonal apertures. An estimate of the average underlying
continuum flux in the same apertures was measured on a continuum map
obtained by averaging fluxes on the two sides of the H$\alpha$ line in
a 19\AA-wide band and used to evaluate the H$\alpha$ equivalent
width of the tails. Fluxes, luminosities (not corrected for internal extinction),
and equivalent widths (EW) are given in Table~\ref{halpha_comp}. It is
worth noting that the tidal tail contributes 31 per cent of the
total H$\alpha$ luminosity of the whole group, and $\sim$ 38 per cent
of this amount (i.e. 12 per cent of the group luminosity in the
emission line) stems from the southern end of the main tidal tail
where TDG7+8 is located. This object and Galaxy 1 have the highest EW(H$\alpha$),
275 and 141 \AA, respectively. All other objects show relatively low equivalent widths.

\subsection{Star formation history of the TDG candidates}

Our previous analysis of optical colors \citep{t03b} indicated a young age (7 to 20 Myr) for the burst
of star formation in the TDG candidates embedded in the tidal tail.
In the attempt to obtain a better estimate of the burst age, we now 
take into account 
the additional information given by long-slit and integral field spectra and we
compare our observations with {\sc starburst99} evolutionary synthesis models \citep{lei99}.
This is a qualitative comparison, not a fitting procedure and 
is only meant to provide a rough indication of the burst age of the TDG candidates,
useful to our understanding of the evolution of the system. Modelling
the star-forming clumps would require that the presence of the underlying older stellar 
populations stemming from the parent galaxy be taken into account, since it contributes to 
the observed spectra. This is beyond the scope of this paper.
For the comparison of optical colors with the models we use 
our measured magnitudes after subtraction of the contribution of the tidal tail \citep{t03b}.

Total spectra of the objects obtained with integral field data (Section 2.2.3) have too small a 
spectral range to be compared with synthetic spectra. However they are useful for 
estimating the H$\alpha$ equivalent width of the objects (Table~\ref{emlines_tot}) and to
determine a scale factor for slit-spectra to compensate for limited spatial coverage.
While the total flux of the continuum of TDG 7+8 is not significantly different from that 
measured on slit spectra, for TDG 10 and TDG 3+9 we measured fluxes a factor of $\sim$ 2.0 and 1.6
higher than in the slit-spectra taken at positions 10 and 9, respectively.
A correction for internal extinction was applied before comparison with the synthetic spectra.
This proved particularly difficult for TDG 7+8 because of the patchy distribution of dust
across it (see Section 3.2). For this object we chose to compare the models with both the 
spectrum obtained at the ESO 3.6-m  telescope and that obtained at the Magellan telescope. 
The first is the sum of the spectra at positions 7 and 8, and has been corrected for extinction
using the average E(B$-$V) = 0.28 mag; the latter includes only object 7 and its extinction 
correction was based on the average value measured along the slit, E(B$-$V) = 0.15 mag.
The two dereddened spectra are very similar, apart from a scale factor in luminosity.
In addition, there appears to be a small metallicity gradient across the regions sampled 
by the two spectra. 

For TDG 3+9 we used the slit-spectrum at the position 9 (which has a better signal-to-noise ratio
than that at the position 3), corrected for the relevant E(B$-$V) and scaled in intensity to match
the total continuum flux measured on the integral field data. The spectrum of TDG 10 was prepared
in a similar way.

For comparison, we chose 
models\footnote{We used the dataset available at http://www.stsci.edu/science/starburst99/} 
simulating an instantaneous burst with a Salpeter initial mass function between masses 
M$_{\mathrm l}$ = 1 M$_{\odot}$ and M$_{\mathrm u}$ = 100 M$_{\odot}$ and metallicities 
Z = 0.004 ($\sim$ 0.2 Z$_{\odot}$) and Z= 0.008 ($\sim$ 0.4 Z$_{\odot}$), thus approaching 
those measured in our spectra (Z = 0.2 - 0.3 Z$_{\odot}$).
Models, calculated for a total mass of 10$^6$ M$_{\odot}$ and including the nebular continuum, 
have been scaled to match
the observed, extinction-corrected luminosity. Since the luminosity scales with the mass,
the comparison with the models gives us a rough idea of the mass of the TDG candidates.

\begin{figure*}
\includegraphics[width=\textwidth]{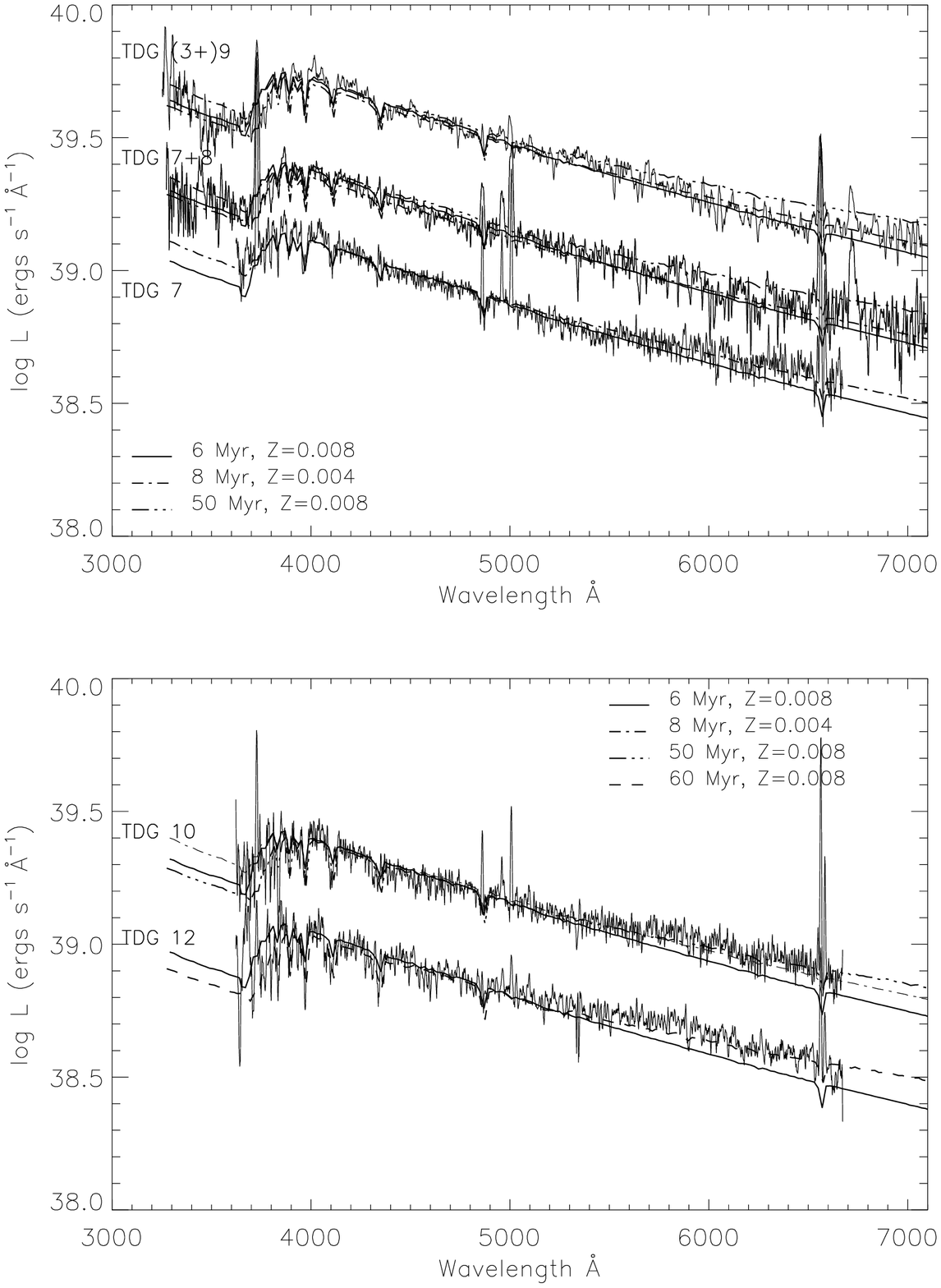}
\caption{Observed, dereddened spectra of the TDG-candidates (see text) overlapped with {\sc starburst99} instantaneous 
burst models. The models assume a Salpeter IMF with masses between 1 and 100 M$_{\odot}$ and are
calculated for a total mass of 10$^6$ M$_{\odot}$. On this plot they are scaled to match the
intensity of the observed spectra. The masses implied for the TDG-candidates are in the range 
1 to 2$\times$10$^7$ M$_{\odot}$. An offset of $-$0.2 has been applied to TDG 12 for better 
visibility. }\label{tdg_models}
\end{figure*}

We find that measured H$\alpha$ equivalent widths are in agreement with models calculated
for burst ages between 5.5 and 8.5 Myr, with the exception of TDG 12, whose lower EW(H$\alpha$)
is consistent with an age of $\sim$ 10 Myr.
Optical colors, after correction for internal extinction, are in rough agreement with the same
age interval.
Out of the spectral energy distributions calculated with the above constraints, those showing
reasonable agreement with the spectra of TDG 7+8 and TDG 3+9 are the models for ages of 6 - 8 Myr
and metallicities Z = 0.004 - 0.008.
The models are overplotted on spectra in Fig.~\ref{tdg_models}.
In particular, TDG 7+8 (and especially the spectrum at the position 7) appears to be in good
agreement with the model at lower metallicity and age 8 Myr, for a total mass of 
1.8$\times$10$^7$ M$_{\odot}$. 
TDG 3+9 is in better agreement with the higher metallicity 6 Myr model and a mass of
2.3$\times$10$^7$ M$_{\odot}$.
The same 8 Myr model reproduces the
continuum of TDG 10 reasonably well (with mass 2.0$\times$10$^7$ M$_{\odot}$). However this object 
shows deeper Balmer absorption lines than the model, suggesting a non-negligible contribution by an 
older stellar population. In fact, a 50 Myr model better reproduces the continuum in the red part of 
the spectrum and the Balmer absorption lines (although not the continuum at the blue end of the range,
nor the H$\alpha$ equivalent width).

Finally, the continuum and the strong Balmer absorption lines of TDG 12 are in good agreement with a 60 Myr
synthetic spectrum. The scale factor in luminosity with respect to the model would imply 
a mass of 1.1$\times$10$^8$ M$_{\odot}$, an unlikely high value in our opinion.
The EW(H$\alpha$) for this model is far lower than that observed.
The comparison process is made even more difficult by the higher uncertainty on
the internal extinction of this clump with respect to the others.
We speculate that, in this case, there might also be a superposition of a relatively young and old stellar populations.
We note that the TDGs in the central part of the tail (objects 10 and 12) are those with the highest
contribution to the optical broad band fluxes from the tidal tail \citep{t03b}.

To summarise, most clumps have properties suggesting ages of their most recent burst of star formation 
in the range 6 to 8 Myr with total masses of $\sim$ 2$\times$10$^7$ M$_{\odot}$.

\section{LIGHT CONTRIBUTED BY TIDAL MATERIAL}

\begin{figure*}
\includegraphics[width=\textwidth]{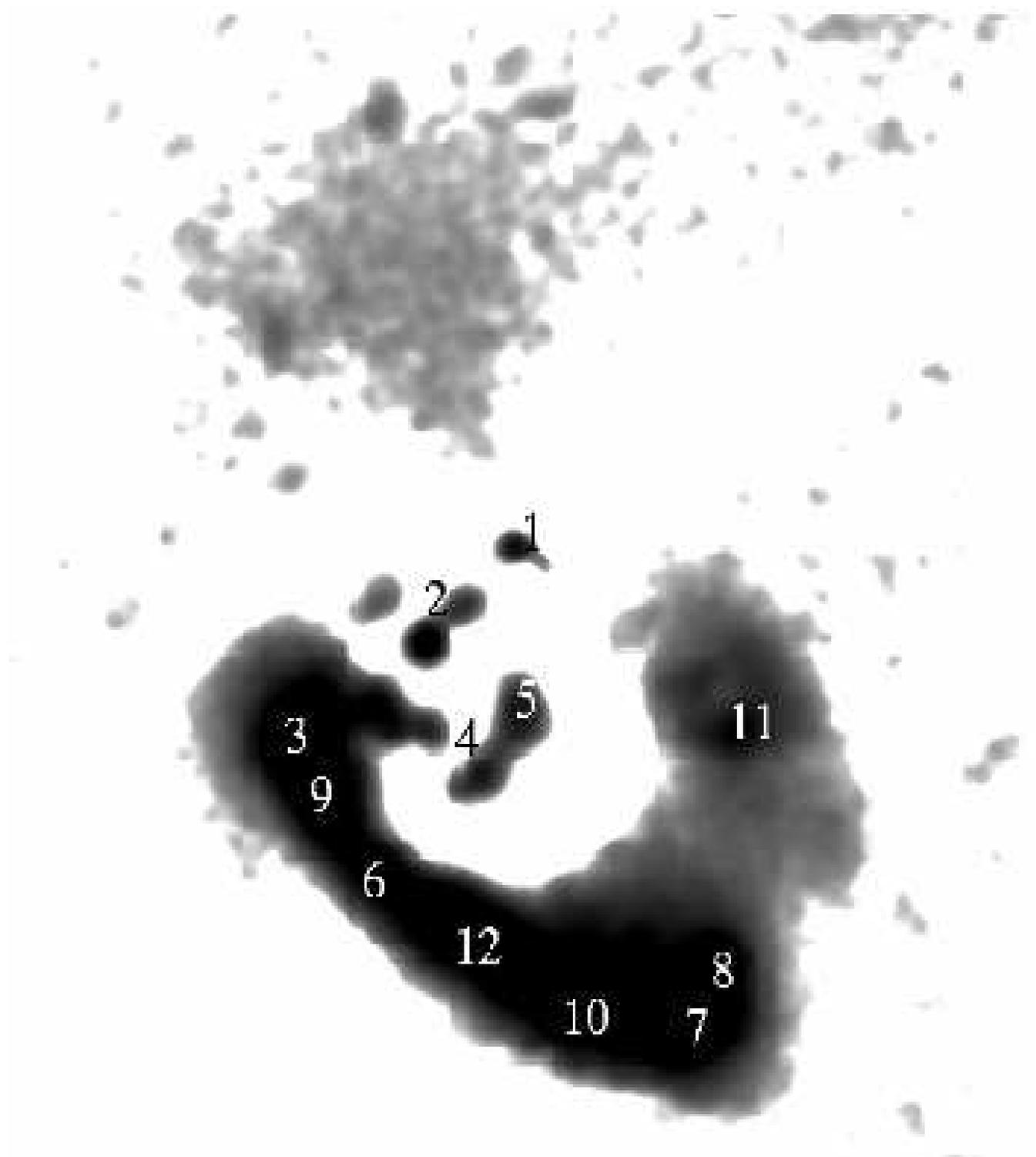}
\caption{ ESO 3.6-m star- and galaxy-subtracted B band image of the tidal features/diffuse light
of CG J1720-67.8. Only features brighter than 26.6 mag arcsec$^{-2}$ (4$\sigma$) are
shown. Object positions are marked according to Fig.~\ref{array_pos}.}\label{diff_light}
\end{figure*}

As noticed above, the main tidal tail considerably contributes to the H$\alpha$ luminosity
of the group. From our optical broad-band images 
\citep[ESO 3.6 m telescope $BVR$-exposures, see ][for a full description of the data and the
reduction steps]{t03a} we are also able to estimate the contribution
of this tidal tail and other tidal material/ faint diffuse light to the total luminosity of the
system. For this purpose the best-fitting galaxy models (obtained as described in \citet{t03a}) 
were subtracted from the optical images and foreground stars were subtracted by means of a 
point-spread-function fitting procedure with {\sc daophot} \citep{sh88}. 
A bi-dimensional fit to the background was also subtracted, and
the images were edited to mask saturated stars and residuals left over by the star-subtraction 
procedure. The $B$ and $R$-band images were convolved with a Gaussian to match the seeing of
the $V$ band image. The resulting images reached a surface brightness limit (3 $\sigma$ level) 
of 26.9 mag arcsec$^{-2}$ in $B$ and 26.0 mag arcsec$^{-2}$ in $R$. A smoothing of the $V$-band
image with a 3-pixel box-car median filter allowed us to reach a sensitivity
similar to the other bands, with a 3$\sigma$ brightness limit of 26.3 mag arcsec$^{-2}$.
The diffuse and tidal light distribution is marginally affected by two bright stars
located at the boundaries of the features of interest. These regions were masked. Isophotal contours are more uncertain in the vicinity of the masked
regions, but measurement of the total light of the tidal features is not significantly 
affected. According to Poissonian statistics, and taking into account the smoothing, we estimated
errors of order $\la$ 0.01 mag. As in the case of galaxy magnitudes, the calibration error
($\la$ 0.05 mag) is dominant.
We find that the tidal material as a whole represents about 31, 27 and 25 per cent of the
total luminosity of the system in $B$, $V$, and $R$, respectively.
The material, including what was previously indicated as an irregular halo of diffuse
light, appears mainly organized in three substructures, namely i) the arc-shaped prominent tidal
tail, ii) a fainter tail-like feature including object 11 and extended from north-west to south,
reaching (or possibly crossing in projection) the southern part of the brightest tail, and 
iii) some diffuse light arranged in a cone-like shape extending north-ward with the apex next 
to the center of Galaxy 1. The latter structure is detected only at 3$\sigma$-level in the $V$
image, while is visible above 4$\sigma$ in the $B$ and $R$ images. 
Fig. ~\ref{diff_light} shows the structures brighter than 4$\sigma$ (26.6 mag arcsec$^{-2}$)
in the $B$-band; residual light from galaxy-model
subtraction is also visible, and has been taken into account when calculating the total
luminosity of the system. The highest contribution to the total luminosity is given by the
bright tidal tail. The total magnitudes of the three tidal features (measured within the
4$\sigma$-level isophote) and their percentage
contribution to the total luminosity of the galaxy group are listed in Table~\ref{tails}. We
notice that the two tails contribute to the total light in slightly decreasing percentage from
$B$ to $R$, while the opposite is true for the cone-like feature, which shows also a widening in
the $V$ and $R$ images. The dominance of an older stellar population in this last component 
might provide an explanation.

\section{EVOLUTIONARY STAGE OF CG J1720-67.8}

\subsection{Possible encounter geometries}

Tails and bridges in interacting systems commonly have a gravitational origin, as was demonstrated
by \citet{tt72}. The kinematics of tidal tails can be used to infer the approximate spin 
geometry of a galaxy encounter \citep[e.g.][]{hvg96,mb98}. 
Here we use the information from the H$\alpha$ velocity field of CG J1720-67.8 together with
the other observational results to speculate about the encounter geometry of its members.
In the following we summarise and discuss the kinematic and star-formation properties of
the galaxy group and propose a few possible scenarios for the interaction history that gave
it its present configuration. Our interpretation is also based on the comparison with observations
and numerical modelling of other merging systems from the literature, however 
adequate numerical simulations specially devoted to model this system are necessary to establish
the validity of any of the suggested scenarios.
We postpone the comparison with detailed simulations to a future work.

The appearance of CG J1720-67.8 is dominated by a bright tidal tail in which several
star-forming clumps are embedded. 
The continuous kinematic trend along this tidal tail across the clumps (Fig.~~\ref{mosaic_hevel}) 
suggests that these clumps are indeed associated with the tail and not projected on to it.
Some systems show signs of return of tidal material to the progenitor galaxy's main body
\citep[e.g. NGC 7252, ][]{hi94} and numerical simulations have confirmed this fate for the
tidal material more tightly bound to the progenitor \citep[e.g.][]{hm95,bh96}.
In our data there is no evidence of return of tidal material.  
In fact the tidal tail is estimated to be younger than 200 Myr and 
most likely even the bound tidal material closest to the galaxy's main body has not yet inverted 
its motion. 
The velocity field of the group shows a global trend, with the kinematic axis coincident
with that of Galaxy 4. However the global kinematic centre has a $\sim$ 5.9 arcsec offset
with respect to the centre of this galaxy. 
We interpret this offset as an effect of the interaction with one of (or both) the companion
galaxies. The asymmetry of the velocity curve of Galaxy 4 at radii $\geq$ 3 arcsec supports the 
idea that the kinematics are influenced by the interaction.

Galaxy 1 has an anomalous velocity field and shows little rotation, if any. Its recessional velocity is
comparable to the systemic velocity of Galaxy 4. 
We note that a bridge emanates from Galaxy 1 toward Galaxy 2 (in projection) and, according to the
velocity field, a bridge of ionized gas, at systemic velocity, seems to connect Galaxies 1 and 4.
 
In prograde major mergers between disc galaxies the formation of two opposite tidal tails
(one per galaxy) is expected, while in retrograde mergers the formation of well defined tails
would be suppressed in favour of plumes \citep[as shown in ][]{tt72,w79}. 
A strong response in the disc, hence
the formation of long and robust tidal tails, requires a small impact parameter, although not too
small otherwise the merging would occur before long tails had the time to develop \citep{b92}.
Since tails are formed only after pericentre passage, the galaxies in CG J1720-67.8 must have
already passed the pericentre of their orbits at least once.
The fact that CG J1720-67.8 has only one long, bright tidal tail, without a strong counter-tail, and 
exhibits a second faint tail and a cone-like plume suggests that either a prograde-retrograde encounter,
with small impact parameter, has taken place -- i.e. only one of the galaxies had the spin vector
(roughly) aligned with the orbital angular momentum vector -- or one of the interacting galaxies was
gas-poor.
Since the strong tail is actively forming stars -- it hosts blue clumps whose latest bursts of
star formation are likely to have occurred 6 to 8 Myr ago and it is responsible for 31 per cent 
of the H$\alpha$ luminosity of the group -- it must contain a considerable amount 
of gas, hence it has likely been produced by a gas-rich galaxy. The most likely progenitor is
Galaxy 4, the spiral galaxy to which the tail appears to be connected. As a consequence, we can assume
that the direction of the orbital angular momentum vector is concordant with the spin vector of
Galaxy 4. Since the tail is quite broad, our observational viewpoint is likely to be approximately above
the orbital plane.

Which role did Galaxies 1 and 2 play in this encounter? Were they both involved in the interaction
that produced the strong tail, and if not, which one is the close interacting companion of Galaxy 4?
We try to address these questions by considering the star formation activity and its distribution 
in the group.

The H$\alpha$ emission indicates that star formation activity is present all across the bodies
of Galaxies 1 and 4 (although it peaks in their central parts), in the centre of Galaxy 2,
all along the strongest tidal tail, in the bridges (at low level), and at least in one clump in
the faint tail. We do not have information about star formation in the plume, however this 
feature is redder than the other tidal debris. The emission in the continuum at 1.4 GHz, which is
a good tracer of star formation and in contrast to H$\alpha$ is not affected by extinction, peaks at
the position of Galaxy 4 and extends to the other two galaxies and to the tail (with a secondary
peak at its southern end).
A number of numerical simulations \citep[e.g., ][]{n88,bh91,bh96,mbr93,mh94} have shown that merger events
drive large quantities of gas from the disc to the centre of galaxies inducing a central burst of
star formation. Such a process could account for the star formation at the centre of the early-type
Galaxy 2 which appears to be a dying merger-induced starburst.
\citet{mrb92} found that star formation rates in merging galaxies can be increased by an order
of magnitude for several times 10$^8$ years with most star formation activity located in the central region
of the merging galaxies. In their close-encounter simulations, both prograde and retrograde mergers
lead to strong bursts of star formation. In retrograde mergers the burst begins in the individual 
discs while they are still separated and continues as the gas falls into the central regions of the
merger. Such a situation could be representative of Galaxies 1 and 4.
The high velocity dispersion of the gas observed at the base of the tidal tail and 
in some parts of the galaxies in connection with regions of star formation (Section 2.2.1), might 
indicate that collisionally triggered star formation is taking place. Collisions in the inter-stellar medium, of
interacting galaxies have been indicated by \citet{mbr93} as a second trigger of star formation in interacting
galaxies, in addition to the increased gas density.
On the basis of the above description regarding the tidal tails and bridges, the kinematics and the star 
formation, we argue that Galaxy 1 is likely to be the retrograde interacting companion of Galaxy 4.
This view agrees with the results of evolutionary synthesis modelling of the galaxies \citep{tf03},
which indicate an age of $\sim$ 180 Myr for the interaction-induced bursts in both galaxies,
consistent with estimated age of the strong tidal tail.
In this case the cone-like plume could have originated from Galaxy 1.
Assuming that the R-band luminosity of the two galaxies roughly traces their masses, this would
be a minor merger, with an approximate mass ratio 4:1 for Galaxy 4: Galaxy 1.

As an alternative possibility, the second faint tail and/or the plume are the result of
a previous encounter possibly involving Galaxy 2. No active star-formation is detected in
these fainter tidal structures, except for the presence of a ring-like feature encompassing
the blue clump no. 11 with no detected H$\alpha$ emission. 
Since Galaxy 2 has a de Vaucouleurs bulge and a weak disc component and exhibits active
star formation in its central parts (although at low level), we suggest the 
possibility that Galaxy 2 is the remnant of a relatively recent merger. If this were the case, 
the central star formation could be interpreted as the remains of a burst triggered during the 
merging phase. The faint tail and plumes could be the tidal debris of such a merger, most 
probably involving two relatively gas-poor galaxies (e.g. two early-type spirals).  

To summarise, we suggest the following possible interaction scenarios that could lead to the 
present configuration of CG J1720-67.8:
\begin{enumerate}
\item Galaxy 2 is a merger remnant, responsible for the faint tail and plume.
   This merger is in excess of 500 Myr old, since the bulge of the galaxy has
   had time to relax to a de Vaucouleurs profile and the merger-induced burst 
   of star formation has already faded. 
   Galaxies 1 and 4 have recently undergone a close passage, which has produced bursts of star
   formation in both galaxies and the formation of the prominent blue tidal tail 
   as well as a bridge departing from Galaxy 4 toward Galaxy 1.
   This view is additionally supported by the fact that the two galaxies have similar
   systemic velocities and both of them exhibit evidence of a recent episode of star
   formation involving a big fraction of their bodies. 
   An interaction of both galaxies with Galaxy 2 might be underway, as suggested by the
   presence of the bridge of matter departing from Galaxy 1 toward Galaxy 2  and the
   possible bridge between Galaxies 2 and 4.
   
\item Galaxy 2 was an early-type galaxy to start with and has recently undergone a close
   interaction with Galaxies 1 and 4. As a consequence of this interaction, Galaxy 4 launched
   a strong tidal tail and a bridge, Galaxy 1 launched a faint tail (retrograde encounter?) 
   and a small bridge, and Galaxy 2 gave origin to the cone-like plume extending to the north.
   The encounter drove some gas from the disc to the center of the gas-poor Galaxy 2, which 
   started a small episode of star formation, while stronger episodes of star formation
   were triggered in the gas-rich Galaxies 1 and 4.
   
\end{enumerate}   

The information derived from our observations is not sufficient to unambiguously 
establish the interaction history of the system. 
The above scenarios cannot be proven without encounter simulations specifically suited to 
reproduce the observed geometry of the galaxy group. Hence, at this level, they
remain hypotheses
that require verifying. From observational data alone, we cannot yet establish what the
future outcome of this system will be, although the small separation of the galaxies in space and 
the strong interactions they are undergoing suggest that the group will merge in a
relatively short time.

\subsection{Kinematics of TDG candidates}

As discussed above, the velocity field of the group shows a regular gradient along the tidal 
tail. No particular kinematic structures are detected at the position of the TDG candidates,
except for the strong gradient at the position of TDG3+9, already reported in \citet{t03b}.
However, local velocity gradients of order of $\approx$ 20 km s$^{-1}$ kpc$^{-1}$
have been found in the long-slit spectrum across the southern part of the tail (Section 3.2).
These results may be reconciled when one considers that the seeing conditions during integral 
field
observations of southern part of the tail were poor and would thus tend to smear out small 
local velocity gradients. Additionally, the H$\alpha$ 
S/N ratio along the tidal tail is quite low (see Fig.~\ref{sn_ratio}), which means that the 
error in velocity measurements is of order of 20 km s$^{-1}$ (Section 2.1), sufficient to hide 
such small velocity gradients. 
Although we still 
lack a proof of self-gravitation for our TDG candidates, we believe that the 
observed local gradients and the kinematic discontinuities 
might suggest distinct kinematics in the individual clumps on top of the 
velocity gradient of the tidal tail.

\subsection{Additional indications of the evolutionary stage}

Indications of the evolutionary stage of a galaxy group can come from the estimate of
its neutral hydrogen content. 
In a homogeneous sub-sample of 48 Hickson Compact Groups \citep[HCGs, ][]{hi82}, \citet{vm} found
a mean H\,{\sc i} deficiency Def$_{\mathrm {H\,I}} = 
\log[\mathrm{M_{H\,I_{pred}}}] - \log[\mathrm{M_{H\,I_{obs}}}] = 0.40 \pm 0.07$ with respect to the expected
H\,{\sc i} content for the optical luminosities and morphological types of the member galaxies
and proposed an evolutionary scenario in which the amount of detected H\,{\sc i} would decrease
with evolution by continuous tidal stripping and/or heating. However, for the triplets in the HCG sample,
they found that the H\,{\sc i} deficiency is not significant (Def$_{\mathrm {H\,I}}$ = 0.19 $\pm$ 0.10).
They argued that these triplets could be either unreal or unevolved systems.
We followed the same method \citep{hg84} to determine the Def$_{\mathrm {H\,I}}$ parameter for CG J1720-67.8.
Our H\,{\sc i} observations failed to detect emission
at the position of CG J1720-67.8, so we compare the upper limit M$_{\mathrm {H\,I}}<2.3\times10^9$ M$_{\odot}$,
estimated in Section 4.1, with the expected H\,{\sc i} content of the system as well as with
the amount of H\,{\sc i} found in other well studied interacting/merging systems.
Taking into account the extinction-corrected B-band magnitudes and the morphological types obtained from
our photometric measurements \citet{t03a}, we determined the expected H\,{\sc i} masses for the 
individual galaxies and used their sum as the predicted H\,{\sc i} content of the whole galaxy group,
M$_{\mathrm {H\,I_{pred}}}$ = 4.9 $\times$ 10$^9$ M$_{\odot}$. This yields a lower limit Def$_{\mathrm {H\,I}} \geq 0.3$, i.e. a significant H\,{\sc i} deficiency.

In a study of five members of the Toomre sequence of merging systems, \citet{hvg96} found a tendency for the
H\,{\sc i} content in the galaxy bodies to progressively decrease going from earlier to later stage mergers. 
While in earlier stages the H\,{\sc i} is found mainly associated with the galaxies, in late stage mergers 
the remnant's body is devoided of H\,{\sc i} in favour of the tidal tails.
The putative two-disc merger NGC3256 has been found to have a mass of  H\,{\sc i}-emitting gas 
M$_{\mathrm {H\,I}}$ = 6.2 $\times$ 10$^9$ M$_{\odot}$ with 75 per cent of the mass in the tidal tails
\citep{e03}. In all cases, the H\,{\sc i} masses of the above cited merging systems are higher than
that estimated for CG J1720-67.8.

Considering the total B-band luminosity of CG J1720-67.8 
after correction for Galactic extinction and average internal extinction E($B-V$) = 0.5 mag,
L$_B$ = 1.1$\times$10$^{11}$ L$_{\odot, B}$, we find an upper limit to the H\,{\sc i} mass
per unit blue luminosity M$_{\mathrm{H\,I}}$/L$_B$ $<$ 0.02 M$_{\odot}$ L$_{\odot,B}^{-1}$.
The ratio does not change significantly if the contribution to the B-luminosity of the
early-type Galaxy 2 (whose contribution to the H\,{\sc i} mass is probably negligible) is subtracted.
This value is lower than the lowest value found in the Toomre sequence of mergers 
\citep[0.07 for NGC7252][]{hvg96}. This fact again points to an H\,{\sc i} deficiency in CG J1720-67.8.

The bright tidal tail of CG J1720-67.8 has an H$\alpha$ luminosity, L(H$\alpha$) = 1.1 $\times$ 10$^{41}$
ergs s$^{-1}$, considerably higher than most tidal tails of the merging galaxies in the Toomre sequence 
[L(H$\alpha$+[N\,{\sc ii}]) = 10$^{39}$ to 3.7 $\times$ 10$^{40}$ ergs s$^{-1}$; \citet{hvg96}] and
its percentage contribution to the global H$\alpha$ luminosity ($\sim$ 19 per cent) is comparable to 
that of the knotty northern tail of NGC 4676. Hence CG J1720-67.8, besides being particularly poor in
atomic gas, displays a star formation activity significantly more intense (roughly by an order of magnitude
in the tidal tail and at least a factor of 4 globally) than the merging galaxy pairs we used as a term
of comparison. Additionally, its global star formation rate is roughly twice as high as 
that measured for HCG 31 \citep{ler04}, a strongly interacting compact galaxy group having some
common properties to CG J1720-67.8 and believed to be already very evolved \citep{ru90,am02}. 
We suggest that this star formation activity is, at least partially, responsible
for the H\,{\sc i} deficiency of the galaxy group. Part of the H\,{\sc i} could also have been dispersed
(at low density level) in the group surroundings by the interaction process, thus becoming particularly 
difficult to detect.

\section{CONCLUSIONS}

In this paper and our previous works \citep{wtk99,t03a,t03b} we have collected a series
of multiwavelength observations with the main aim of establishing the evolutionary state
of the ultra-compact galaxy group CG J1720-67.8. Our studies have revealed this system
as an interesting candidate to represent the rarely observed phase in the
evolution of compact groups that precedes the final merging.
In particular in this work we have considered the kinematics of the group components -- based
on both integral field and long slit optical spectroscopy -- and the
intensity and distribution of star formation activity -- based on H$\alpha$ and radio continuum
observations -- together with the photometric properties of the
whole group and its tidal debris to put some constraints on the interaction history of
CG J1720-67.8. We use encounter simulations in the literature as well as 
examples of merging galaxy pairs to interpret our observations and outline some possible 
evolutionary scenarios. However, the presence of tails and bridges involving all three galaxies
suggests a more complex interaction than those usually simulated considering only galaxy pairs.
In CG J1720-67.8 we have most probably a three-body encounter or an even more complicated case
with one of the galaxies being already the outcome of a relatively recent merger.

Although other interpretations are possible, as discussed extensively in Section 7.1, 
we favour a scenario in which Galaxy 4 and Galaxy 1 experienced a prograde-retrograde
close encounter within the last 200 Myr. During this encounter, the strong tidal tail and
the bridge between the two galaxies were formed, and possibly also the cone-like plume
with the apex pointing approximately to the centre of Galaxy 1.
The interaction process triggered strong star formation episodes across the bodies and in the
centres of both galaxies. More recent star formation events ($<$ 10 Myr) have taken place
in the condensations of gas and stars that formed under the action of self-gravity within the
tidal tail. These condensations have properties consistent with tidal dwarf galaxies
in the process of formation. Although these objects cannot be confirmed as real TDGs, they do show
possible signs of self-gravitation.

The role of Galaxy 2 in this recent encounter is not clear. The small bridge departing from
Galaxy 1 in its direction suggests that this galaxy was involved in the interaction.
The presence of a further bridge between Galaxies 2 and 4 is possible but cannot be established 
with certainty because of the significant overlap between the two galaxies in projection on the sky.
Our new data have confirmed the presence of low level star formation concentrated in the central part
of Galaxy 2. This fact, together with the surface brightness distribution indicating a dominant
de Vaucouleurs bulge, suggested the interesting hypothesis that Galaxy 2 is a merger remnant
still showing the remnant of a central burst of star formation triggered by the merger event.
If this was the case, the faint secondary tail seen on the western side of the group could have
emanated from Galaxy 2's progenitors. This is also an alternative origin for the northern 
cone-like plume.
Since our data do not have sufficient spatial resolution to disentangle any possible double nucleus
or other substructure within Galaxy 2, the merger hypothesis remains speculative.
The proximity of the three galaxies in space and their strong interactions suggests that 
they will merge in a relatively short time.
Furthermore, several indications of significant  H\,{\sc i} deficiency in the group (Section 7.3),
according to the evolutionary scenario proposed by \citet{vm}, support the idea that the galaxy system 
is already evolved.
Dedicated dynamical modelling is necessary to evaluate the validity of any of the proposed
scenarios and predict the future evolution of the system.

\paragraph*{Acknowledgments}
ST acknowledges support by the Austrian Science Fund (FWF) under project 
no. P15065 and is grateful to the staff of the Anglo-Australian Observatory for their
helpful assistance during the observations at the AAT, and to the Australia Telescope National 
Facility (Epping, NSW) for hospitality during the preparation of part of this work. 
ST would like to thank S. Ciroi for useful suggestions and for making available, 
together with M. Radovich, their scripts for integral field data handling.

\clearpage

\clearpage

\clearpage

\clearpage

\clearpage

\clearpage

\clearpage

\clearpage

\clearpage

\clearpage

\begin{table}
\caption{Emission-line measurements on total spectra$^{\mathrm{a}}$}\label{emlines_tot}
\begin{tabular}{lllll}
\hline
\hline
Object &  F(H$\alpha$) & log($[$N\,{\sc ii}$]\lambda\, 6583$/H$\alpha$) & log($[$S\,{\sc
ii}$]\lambda\, 6716,6731$/H$\alpha$) & EW(H$\alpha$)\\
       \hline
 & & & & \\
G1     &  243.3  $\pm$ 2.5  &  -0.75 $\pm$ 0.03& -0.66 $\pm$ 0.05 & 141\\
G2     &   75.2  $\pm$ 11.3 &  -0.38 $\pm$ 0.26& -0.23 $\pm$ 0.25 & 9\\
G4     &  248.5  $\pm$ 4.1  &  -0.43 $\pm$ 0.03& -0.34 $\pm$ 0.05 & 52\\
TDG3+9 &   38.9  $\pm$ 2.7  &  -0.51 $\pm$ 0.14& -0.55 $\pm$ 0.27 & 40\\
TDG7+8  &  89.8  $\pm$ 3.3  &  -0.79 $\pm$ 0.10& -0.57 $\pm$ 0.12 & 275\\
TDG 10 &   30.7  $\pm$ 1.7  &  -0.55 $\pm$ 0.09& -0.45 $\pm$ 0.17 & 47\\
TDG 12 &   13.3  $\pm$ 2.7  & ...    & ... & 14\\
\hline
\end{tabular}
\begin{list}{}{}
\item[$^{\mathrm{a}}$] All fluxes are in units of 10$^{-16}$ ergs s$^{-1}$ cm$^{-2}$. Equivalent
widths (EW) are in \AA.  A correction for Galactic extinction has been applied.
\end{list}
\end{table}

\begin{table}
\caption{Observing parameters for the radio observations}\label{radio-obslog}
\begin{tabular}{lcccc}
\hline
\hline
 & & & & \\
Date   &               2002 Jan 17 & 2002 Feb 25 & 2003 Aug 4 & 2003 Aug 5 \\
Array  &                  750A     &   1.5A      &   6D       & 6D \\
\multicolumn{5}{c}{HI Observations}\\
Frequency (MHz)    &      1359   &      1359  & & \\
Bandwidth (MHz)    &         8   &         8  & &  \\
No. channels       &       512   &       512  & & \\
Observing time (hrs) &      12   &        12  & & \\
 (at each frequency)& & & & \\
\multicolumn{5}{c}{Continuum observations}\\
Frequency          &      1384   &      1384  &      4800   &    1384 \\
                    &            &            &      8640   &    2368 \\
                    &            &            &             &    5184 \\
                    &            &            &             &    5952 \\
Bandwidth           &      128   &       128  &       128   &     128 \\
No. channels        &       32   &        32  &        32   &      32 \\
Observing time (hrs) &      12   &        12  &        12   &       5.5 \\
 (at each frequency)& & & & \\
\hline
\end{tabular}
\end{table}

\begin{table}
\caption{Results of the HI observations} \label{HIobs}
\begin{tabular}{lc}
\hline
\hline
 & \\
Weighting          &            Natural \\
Beam FWHP          &            44$^{\prime\prime}$x39$^{\prime\prime}$ \\
RMS (mJy/beam)     &             1.5 \\
Velocity range (cz) (km s$^{-1}$)&  12800 - 13800 \\
Flux upper limit (Jy km s$^{-1}$) &     0.3 \\
 (3-$\sigma$; 200 km s$^{-1}$) & \\
HI mass (M$_{\odot}$)  &   $<2.3\times10^9$\\
\hline
\end{tabular}
\end{table}

\clearpage 

\begin{table}
\caption{Results of the radio continuum observations}\label{cont-obs}
\begin{tabular}{lllll}
\hline
\hline
 & & & & \\
Frequency (MHz)        &          1384   &   2368   &   5312$^{\mathrm{a}}$ & 8640 \\
Weighting              &         Uniform & Natural  & Natural  & Natural \\
Beam FWHP              &        6\farcs5$\times$6\farcs3 &7\farcs0$\times$5\farcs4 &2\farcs9$\times$2\farcs5&1\farcs7$\times$1\farcs4 \\
RMS ($\mu$Jy/beam)     &           42     &   75     &    26    &    40 \\
Peak flux density (mJy/beam) &    1.5     &  0.71    &   0.43    &  0.20 \\
RA(J2000) at peak      &       17 20 28.8 & 17 20 28.7&  17 20 28.8 & 17 20 28.9 \\
Dec(J2000) at peak      &     -67 46 31   & -67 46 32 & -67 46 31 & -67 46 31 \\
Position error           &        1\farcs0    &   1\farcs0    &   0\farcs5    &  1\farcs0 \\
Total flux density$^{\mathrm{b}}$ (mJy) &    4.2     &   ...     &  $>$1.1   &    ... \\
\hline
\end{tabular}
\begin{list}{}{}
\item[$^{\mathrm{a}}$]Mean of the three observing frequencies 4800, 5192 and 5952 MHz.\\ 
The frequencies were chosen to maximise coverage in the uv-plane.
\item[$^{\mathrm{b}}$]Object too extended for accurate total flux measurement above 
1384 MHz.
\end{list}
\end{table}

\begin{table}
\caption{Polygonal aperture H$\alpha$ measurements of the group components}\label{halpha_comp}
\begin{tabular}{llll}
\hline
\hline
Object & F(H$\alpha$) & L(H$\alpha$) & EW(H$\alpha$)$^{\mathrm{a}}$ \\
       & (ergs s$^{-1}$ cm$^{-2}$) & (ergs s$^{-1}$) & \AA\ \\
       \hline
 & & & \\
G1         &   2.82\,10$^{-14}$ & 1.10\,10$^{41}$ & ... \\
G2         &   4.04\,10$^{-15}$ & 1.60\,10$^{40}$ & ...  \\
G4         &   2.59\,10$^{-14}$ & 1.01\,10$^{41}$ & ... \\
Main tail  &   2.93\,10$^{-14}$ & 1.14\,10$^{41}$ & 28.2 \\
Secondary tail & 4.11\,10$^{-15}$ & 1.60\,10$^{40}$ & 15.9\\
TDG7+8     &   1.10\,10$^{-14}$ & 4.3\,10$^{40}$ & ... \\
\hline
\end{tabular}
\begin{list}{}{}
\item[$^{\mathrm{a}}$] For the EW of individual galaxies see Table~\ref{emlines_tot}. 
\end{list}
\end{table}

\newpage

\begin{table}
\caption{Optical luminosity$^{\mathrm{a}}$ of tails and diffuse light}\label{tails}
\begin{tabular}{lllllll}
\hline
\hline
Object & $B$ & $V$ & $R$ & L$_B$ & L$_V$ & L$_R$ \\
       & (mag) & (mag) & (mag) & (L$_{\odot,B}$) & (L$_{\odot,V}$) & (L$_{\odot,R}$) \\
       \hline
 & & & & & & \\
Main tail  & 16.9  & 16.4 &  16.1 & 8.8\,10$^9$ &  7.7\,10$^9$ &  6.3\,10$^9$ \\
Percentage of total luminosity & 25.5 & 21.3 & 18.9  & ... & ... & ... \\
Faint tail  & 19.0  & 18.5 &  18.3  & 1.3\,10$^9$  & 1.1\,10$^9$  &  8.2\,10$^8$ \\
Percentage of total luminosity & 3.6 & 3.2 & 2.7 & ... & ... & ... \\
Cone-like feature & 19.4  & 18.6 & 18.0 &  8.8\,10$^8$ &  1.0\,10$^9$ &  1.1\,10$^9$ \\ 
Percentage of total luminosity & 2.3 & 2.9 & 3.4  & ... & ... & ... \\
Combined tidal features  &  16.6  & 16.1 & 15.8  &  1.2\,10$^{10}$ & 1.0\,10$^{10}$  &  8.2\,10$^9$ \\
Percentage of total luminosity & 31 & 27 & 25  & ... & ... & ... \\
\hline
\end{tabular}
\begin{list}{}{}
\item[$^{\mathrm{a}}$] As solar values we used B$_{\odot}$ = 5.48 mag, V$_{\odot}$ = 4.83 mag, 
and R$_{\odot}$ = 4.31 mag.
\end{list}
\end{table}

\end{document}